\documentclass[final,5p,times,twocolumn,authoryear]{elsarticle}

\usepackage{amssymb}

\usepackage[table,xcdraw]{xcolor}
\usepackage[normalem]{ulem}
\usepackage{array}
\newcolumntype{x}[1]{>{\centering\arraybackslash\hspace{0pt}}p{#1}}

\definecolor{ao}{rgb}{0.0, 0.5, 0.0}

\newcommand{\black}{\color{black}}

\usepackage{enumitem}
\usepackage{multirow}
\usepackage{array}
\usepackage{tcolorbox}
\usepackage{wasysym}

\newcommand{\rot}[1]{\rotatebox{90}{#1}}
\let\origcheckmark\checkmark
\renewcommand{\checkmark}{\scalebox{0.8}{\origcheckmark}}
\newcommand{\smallfullmoon}{\scalebox{0.7}{\fullmoon}}
\newcommand{\smallnewmoon}{\scalebox{0.7}{\newmoon}}

\setlist[description]{leftmargin=1em}

\usepackage[T1]{fontenc}

\newtcolorbox{fancybox}[1][]{
  colback=gray!10,
  colframe=white,
  boxrule=0pt,
  sharp corners,
  width=\linewidth,
  before skip=2pt,
  after skip=2pt,
  boxsep=0pt,
  #1
}

\PassOptionsToPackage{hyphens}{url}
\usepackage[hidelinks]{hyperref}
\hypersetup{
    colorlinks,
    linkcolor=blue,
    citecolor=blue,
    urlcolor=blue
}
\expandafter\def\expandafter\UrlBreaks\expandafter{\UrlBreaks
    \do\a\do\b\do\c\do\d\do\e\do\f\do\g\do\h\do\i\do\j%
    \do\k\do\l\do\m\do\n\do\o\do\p\do\q\do\r\do\s\do\t%
    \do\u\do\v\do\w\do\x\do\y\do\z\do\A\do\B\do\C\do\D%
    \do\E\do\F\do\G\do\H\do\I\do\J\do\K\do\L\do\M\do\N%
    \do\O\do\P\do\Q\do\R\do\S\do\T\do\U\do\V\do\W\do\X%
    \do\Y\do\Z\do\*\do\-\do\~\do\'\do\"\do\-}%

\newcommand{\urlDate}{last accessed 2021-02-22}
\newcommand{\furl}[1]{\footnote{\url{#1} (\urlDate).}}

\journal{DFRWS USA 2025}

\begin{document}

\begin{frontmatter}



\title{SoK: Timeline based event reconstruction for digital forensics: Terminology, methodology, and current challenges}


\author[label1]{Frank Breitinger\corref{cor1}}
\ead{frank.breitinger@uni-a.de}
\cortext[cor1]{Corresponding author}
\author[label2]{Hudan Studiawan}
\ead{hudan@its.ac.id}
\author[label3]{Chris Hargreaves}
\ead{christopher.hargreaves@cs.ox.ac.uk}

\affiliation[label1]{organization={Institute of Computer Science, University of Augsburg},
             city={Augsburg},
            country={Germany}}

\affiliation[label2]{organization={Department of Informatics, Institut Teknologi Sepuluh Nopember},%
           city={Surabaya},
           country={Indonesia}}

\affiliation[label3]{organization={Department of Computer Science, University of Oxford},
             city={Oxford},
             country={United Kingdom}}

\begin{abstract}

Event reconstruction is a technique that examiners can use to attempt to infer past activities by analyzing digital artifacts. Despite its significance, the field suffers from fragmented research, with studies often focusing narrowly on aspects like timeline creation or tampering detection. This paper addresses the lack of a unified perspective by proposing a comprehensive framework for timeline-based event reconstruction, adapted from traditional forensic science models. We begin by harmonizing existing terminology and presenting a cohesive diagram that clarifies the relationships between key elements of the reconstruction process. Through a comprehensive literature survey, we classify and organize the main challenges, extending the discussion beyond common issues like data volume. Lastly, we highlight recent advancements and propose directions for future research, including specific research gaps. By providing a structured approach, key findings, and a clearer understanding of the underlying challenges, this work aims to strengthen the foundation of digital forensics. 

\end{abstract}


\end{frontmatter}



\section{Introduction}

Event reconstruction involves recreating past events by analyzing digital artifacts, allowing examiners to determine system activities and make informed conclusions about what occurred. 
While traditional forensic science benefits from a well-defined framework summarizing the field \citep{ribaux2023police}, event reconstruction in digital forensics is often discussed in fragmented terms focusing on tasks such as super timeline creation \citep{gudhjonsson2010mastering, metz2024log2timeline}, tampering detection \citep{palmbach2020artifacts, studiawan2021anomaly} or environmental peculiarities \citep{schatz2006correlation}. As a result, research has centered on these narrow aspects, leaving broader challenges underexplored or overlooked.
The absence of a unified perspective has led to a proliferation of terms, making it difficult to discuss event reconstruction comprehensively or find relevant research, e.g., some studies use the term artifact \citep{harichandran2016cufa}, others refer to observable facets \citep{jaquet2021formalized}. Terms such as events \citep{carrier2004defining}, user actions, interactions, or clicks \citep{neasbitt2014clickminer} are inconsistently used in literature.

\emph{The three contributions:} 
First, the article discusses concepts and definitions in timeline-based event reconstruction and integrates them into a new visual model (the timeline-based event reconstruction model or \emph{TER-Model}), divided into four quadrants, integrating digital forensic timeline-based terminology and  \cite{ribaux2014police} model. 
Second, with this delineation, we provide a thorough discussion of the issues associated with timeline-based event reconstruction. These issues can be used to evaluate event reconstructions and identify areas of uncertainty in the results. They can also be used to systematically identify weaknesses in the timeline generation and analysis techniques and contribute to a knowledge base of such weaknesses such as SOLVE-IT \citep{hargreaves2025solve}.
Third, we provide future research directions needed within each quadrant of the event reconstruction process.
This paper is predominantly theoretical, aiming to harmonize timeline-based event reconstruction terminology, however, a practical illustration of the use of the model is available online\footnote{\url{https://github.com/chrishargreaves/TER-model-example}}.

\emph{Not in scope:} The identification of relevant devices (computer profiling, \citet{marrington2007event}), legal constraints or ethical issues \citep{losavio2015cyber}, technical challenges such as encryption, sophistication of crime \citep{karie2015taxonomy}, or very general challenges, e.g., that ``results must be reproducible and verifiable'' \citep{soltani2019formal}.

\emph{Outline:} The next section summarizes core works in event reconstruction which served as a foundation for this work. 
Subsequently, Sec.~\ref{sec:terms-def} presents terms and technology in existing literature and outlines the terminology used in this article. 
A contribution of this work is the TER-model which is developed and described in Sec.~\ref{sec:model}. Using the model, we identified challenges according to the methodology in Sec.~\ref{sec:methodology-for-challenges} and organized the challenges for event reconstruction in the two main sections: \nameref{sec:challenges-in-reconstruction} and \nameref{sec:challenges_tampering}, which are summarized as key findings in Sec. \ref{sec:keyfindings}.  
Considering these, Sec.~\ref{sec:futureResearchDirections} provides a discussion and identifies specific research gaps. The final section concludes the paper.

\section{Event reconstruction}
\label{sec:event_reconstruction}

\citet{lee2001henry} and many others have discussed event reconstruction for physical crime scenes.
\citet{carrier2004defining, carrier2004event} were the first to define it as applied in digital forensics and presented an event-based investigation framework. Their work defines the basic terminology and introduces a formal process model that mirrors physical crime scene investigations but is tailored to the unique aspects of digital evidence. We borrow from this work as discussed in Sec.~\ref{sec:existing_terms}.

\cite{casey2011digital}'s work includes the practicalities of linking evidence to behaviors and motives.  \citeauthor{casey2011digital} emphasizes three core analysis types: 
(1) temporal which helps establish the timeline of events (the focus of this article),
(2) relational which explores the connections between objects, people, and locations, clarifying how different elements of the crime are related, and 
(3) functional which assesses what was possible or impossible, such as determining how a system or tool was used in the crime. 
\citet{chabot2015event} defines terminology based on existing works, outlines challenges, and evaluates existing approaches. However, the authors limit their challenges to the volume of data and data heterogeneity where this article provides a broader discussion. 
Our work complements these existing works by providing a new visual model and a thorough discussion of challenges and future research.

\section{Terminology}
\label{sec:terms-def}

According to \citet{neale2023fool}, there is a lack of harmonization in terms and definitions. 
This section briefly revisits (Sec.~\ref{sec:existing_terms}) and then highlights the terminology we use for this article (Sec.~\ref{sec:proposed_terms}).

\subsection{Terms and terminology in existing literature}\label{sec:existing_terms}
\citet{carrier2004defining} define an event ``as an occurrence that changes the state of one or more objects''. Over time, researchers suggested to differentiate between low-level and high-level events (human-understandable) \citep{hargreaves2012automated, vanini2024clock} or introduced terms such as `activity' \citep{marrington2007event} or `user-browser interaction' and `click' which are used interchangeably by \citet{neasbitt2014clickminer}. 
\citet{chabot2014complete} defines an event as ``a single action occurring at a given time and lasting a certain duration''. 

\cite{jaquet2021formalized} define an event as ``a complete collection of related things that have happened (or are happening) in a World within a specific closed interval of time. [...] The Event can be considered as a whole entity or as a collection of smaller sub-events''. 
Notably, their framework emphasizes the role of traces and introduces several key concepts, including trace, facet, and observable facet. While these terms are well-established in forensic science \citep{ribaux2023police}, they are less common in digital forensics. Therefore, we adopt a different terminology, while drawing conceptual links to their work.

Similarly, the term \emph{artifact} is used with different meanings. For instance, 
\citet{harichandran2016cufa} compares various definitions and concludes properties an artifact should have such as ``artificiality/external force, antecedent temporal relation, and exceptionality''.
\citet{horsman2019raiders} suggests ``a digital object containing data which may describe the past, present or future use or function of a piece of software, application or device for which it is attributable to''. 
\citet{casey2022crowdsourcing} differentiates between atomic artifacts (``a singular unit of interpretable data that can be extracted from a given data source'') and dependable artifacts (``one or more atomic artifacts needed to expose the atomic artifact of interest''). \citet{lyle2022digital} extends the atomic artifact definition by adding ``...that is useful for addressing questions in forensic investigations'', but assessing usefulness is difficult, subjective and may change over time.

\subsection{Terminology used in this article}\label{sec:proposed_terms}

\paragraph{Environments/systems} An environment/system is a computational setting or a software/hardware system that reacts to events such as user actions, API calls, or sensor inputs. Typically, it is one or more  devices such as computers or smartphones but it could also be a virtual machine, network device, or cloud environment. 
For readability, the remainder of this paper uses the term environments instead of environments/systems. Note we use the plural, i.e., environments, considering that changes may be in one or more environments, locally, remotely, or both.
\black

\paragraph{Artifact} 
This article uses \citet{casey2022crowdsourcing} atomic artifact definition: a singular unit of interpretable data that can be extracted from a given data source.
For simplicity, we will only say artifact throughout the paper.  Examples include log files, registry keys, timestamps, or network traffic data.

\paragraph{Event} 
 
Based on \cite{jaquet2021formalized}, an event is ``a complete collection of related things that have happened (or are happening) in a World within a specific closed interval of time.''
These can be treated as a singular entity or decomposed into smaller sub-events and cause environmental changes. 
This broad definition provides the flexibility for an event to be at the resolution of: 
`file was accessed', or 
`Google search was performed', or 
`user account was used to run a program' (consisting of at least two events: user logged in and user executed binary). \black
Events can be triggered internally, e.g., a cron job, or externally, e.g., someone clicking the mouse. 
%
Note that the distinction between event and sub-event is blurred and it is up to the user to define the granularity. For instance, 
\begin{itemize}[noitemsep]
    \item an event is \emph{sending an email} with sub-events such as opening the email client, typing, establishing a connection to the SMTP server, and sending the message, or
    
    \item an event is \emph{establishing a connection to the SMTP server} with sub-events such as performing a DNS lookup, initiating a handshake, and authenticating the user credentials. 
\end{itemize}

\section{Model for event reconstruction}
\label{sec:model}
\begin{figure*}[t]
    \centering
    \includegraphics[width=0.72\textwidth,trim={1cm .5cm 2cm 1cm},clip]{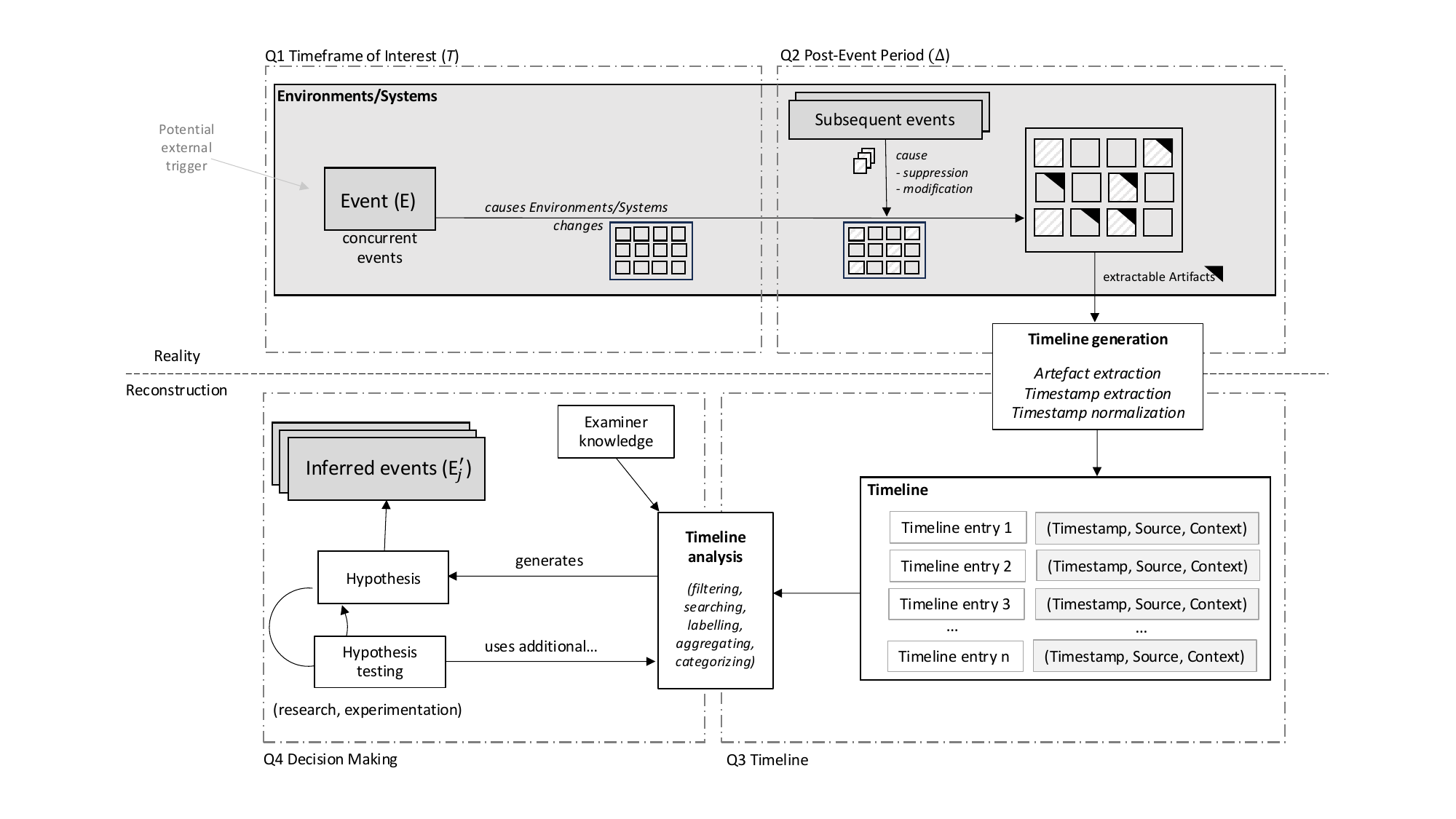}
    \caption{TER-Model: Model of timeline-based event reconstruction in digital crime scenes. The small squares (3x4) in the upper part of the diagram represent changes by the primary event (gray box) and additional changes from subsequent events (white-gray stripes).
    \label{fig:diagram}}
\end{figure*}

This work draws inspiration from \cite{vanini2023presentation}, which, in turn, is influenced by the work of \citet[p226, Fig.~4.4]{ribaux2023police}\footnote{Note, this is an updated version from the previous work by \cite{ribaux2014police} and thus has over a decade of history.}. 
We adjusted these models to align with standard digital forensics terminology and emphasize timeline-based event reconstruction.
Our model, named \emph{TER-Model} (timeline-based event reconstruction), is depicted in Fig.~\ref{fig:diagram} and can be separated into a \emph{reality} space (Sec.~\ref{sec:reality}) and a \emph{reconstruction} space (Sec.~\ref{sec:perception}). Each of these spaces can be further separated resulting in four quadrants (Q1-Q4). Before describing the model, this section first summarizes the goals of temporal event reconstruction which influenced the TER-Model. The summary of systematization of knowledge (SoK) in the TER-Model is shown in Table \ref{tbl:sok}.

\subsection{Goals of temporal event reconstruction}

Temporal event reconstruction aims to accurately recreate the sequence of events that occurred which includes finding gaps and inconsistencies, even if they cannot be accurately filled or corrected. 
Thus, it enables investigators to draw meaningful conclusions about what transpired. 

Event reconstruction involves several interrelated analytical processes that together provide a coherent and defensible narrative of what transpired. At its core is temporal sequencing and correlation, where a precise order of events is created. It may be necessary to analyze their relationships across different timelines to uncover causal links, sequence dependencies, or concurrent activities \citep{adderley2020interactive}. 
Beyond simple chronology, contextual analysis places these events within a broader framework, considering factors such as user behavior, system settings, or external influences to give the data deeper interpretive meaning \citep{chabot2015event}. This groundwork supports hypothesis testing and scenario building, where investigators construct and refine possible explanations for what occurred, evaluating multiple narratives and ruling out those that conflict with the evidence \citep{willassen2008hypothesis,willassen2008finding,batten2012hypothesis}. 
It is crucial that the reconstructed timelines are confirmed through correlation and verification of evidence to ensure consistency and reliability. The goal is to produce a report to support legal proceedings that not only stands up to technical scrutiny, but also serves court proceedings by providing a clear, accurate and accessible story for stakeholders such as lawyers or jurors \citep{chabot2014complete,xu2022visualizing}.

\subsection{Reality and its two dimensions (Q1, Q2)}\label{sec:reality}

\paragraph{Q1: Timeframe of interest $T$} 
This quadrant is an interval that has a start time $t_S$ and an end time $t_E$, i.e., $T = [ t_S, t_E ] $ during which the event ($E$) and sub-events ($e_1, e_2, ... e_m$) occurred. 
Each $E$ or $e$ causes multiple environmental changes, e.g., new log entries, modified registry values, files marked as non-allocated, or updated timestamps.

The event (E) is what we wish to be able to say something about through the event reconstruction process. \cite{carrier2006hypothesis} describes that an event can be any ``an occurrence that changes the state of the system'' and \cite{hargreaves2009assessing} continues that ``digital events occur on a system often as a result of interactions with another digital device, or as a result of interactions with the real world''. However, in \cite{jaquet2021formalized} event is formalized such that these external triggers are integrated into the event itself, defining an event that can capture the very broad, or the very detailed. 
In addition, there are \emph{concurrent events} such as antivirus scanning files resulting in changes not tied to the primary event.

\paragraph{Q2: Post-Event Period ($\Delta$)} 
During this interval $\Delta$, the environment changes caused by $E$ may become intermingled with, altered, or overwritten by an ensemble of other data generated by unrelated \emph{subsequent events}. 
\citet{jaquet2021formalized} categorized these changes as adjunction, suppression, and change.
This second interval ends at time $t_P$ when the data is preserved/extracted, i.e., $\Delta = ( t_E, t_P ]$. {As $t_E$ belongs to $T$, we exclude it here from this interval using a half-open interval.}
It is important to note that not all environment changes can be extracted, such as missing/deleted files or new artifacts without a parser. These gaps may stem from many causes, for example a lack of knowledge in digital forensics, a tool setup, or errors in the timeline generation process. Hence, what can be extracted is named \emph{extractable artifact}, which is therefore context specific.

\paragraph{Timeline Generation} 

Combined with preservation and acquisition, timeline generation bridges the Reality and Reconstruction spaces. \citet{hargreaves2024abstract} define it as a process within a forensic analysis tool for ``extracting timestamps from the file system...[and] applying file specific processing and extracting timestamps from within files such as the Windows Registry, log files, SQLite databases etc., that contain timestamps''. 
This artifact and timestamp extraction is complemented by normalization, which is required since timestamps exist in a variety of formats (e.g., ASCII in a log vs.~little-endian hexadecimal in a proprietary format), and resolutions (i.e., hours, minutes, seconds, nanoseconds, etc.) depending on their source \citep{raghavan2013unitime}. They may also be stored in UTC or local time. Ideally, after normalization, all timestamps should be presented in the same format for better readability and sortability.

\subsection{Perception}\label{sec:perception}

The lower section of the diagram represents how examiners attempt to reconstruct past events 
using reasoning and available evidence. This process involves uncertainty, as the past cannot be revisited, making absolute certainty unattainable.

\paragraph{Q3: Timeline} 
Examiners construct a timeline to facilitate analysis, and the DFPulse 2024 Practitioner Survey \citep{hargreaves2024dfpulse} reports 80.3\% are using timelines `often' or `almost always'. 
Timelines are composed of a series of entries, each derived from individual artifacts that are arranged chronologically. Artifacts may originate from multiple independent data sources, e.g., a computer and a smartwatch.  
While specific implementations store multiple data points per event, fundamentally these \emph{timeline entries} are defined as a 3-tuple $(t, S, C)$:

\begin{itemize}[noitemsep]
    \item The normalized timestamps ($t$) are used to order the timeline chronologically.
    
    \item A source $S$ refers to the specific location from which the timestamp and context originate, such as the Master File Table (MFT), Windows registry, EPROCESS block in memory, or Chrome browser history file. For clarity, $S$ should be as detailed as possible; instead of stating the registry, the exact registry key path should be specified.
    
    \item A context $C$ defines what the timestamp represents, such as the modification timestamp within the Standard Information Attribute (SIA) of MFT entry, or a value in a specific row or field within a database. Given the wide variety of contexts, a generic term is used to encompass the diverse nature of these representations. 
\end{itemize}

These timeline entries should not be conflated with events themselves or `low-level events'  \citep{hargreaves2012automated}.  The context provided by each entry, such as a value in a `modified' or `last change' field within a file system structure, does not inherently represent a specific event, such as a file modification. 
Instead, it reflects environmental behavior that must be understood before making any assumptions about what event occurred. This distinction is critical: while timeline entries provide the raw data needed for event reconstruction, they are not events in and of themselves. Rather, they are normalized, sorted compilations of data that result from parsing artifacts left by events. Therefore, we argue that the term event should be reserved for the inferred actions, while the term timeline entry more accurately describes the data points that examiners use to reach those inferences.

\paragraph{Timeline Analysis}
Timeline analysis bridges Q3 and Q4, and describes the process of moving from having a timeline to reconstructing events, which uses refinement techniques such as:
filtering irrelevant entries, 
highlighting key entries, or 
aggregating entries into more meaningful events \citep{hargreaves2012automated}. 
Several other concepts have been discussed such as event abstraction \citep{studiawan2020automatic,studiawan2023event}, the application of machine learning \citep{khan2006machine}, or visualization \citep{berggren2024timesketch, debinski2019timeline2gui}.
Timeline analysis also draws in \emph{examiner knowledge} to understand potential events that are capable of producing the timeline entries and integrating them into a reasoning process \citep{gladyshev2004finite}.

\paragraph{Q4: Hypotheses and Event Inference} To accurately approach event reconstruction, it is essential to distinguish between the event $E$ that occurred in reality and the inferred event $E'$ which is derived from the analysis of timeline entries. In the context of hypothesis generation, $E'$ represents the best approximation based on the available evidence. We define an inferred event $E'$ as \emph{a reconstructed scenario that may have occurred within a specific time frame, based on the interpretation and analysis of timeline entries and associated artifacts.}
This definition acknowledges the uncertainty in reconstructing past events.

Consideration of the timeline entries in the context of examiner knowledge may result in multiple plausible scenarios \citep{jaquet2021formalized, gladyshev2004finite}. \cite{hargreaves2009assessing} states ``if there are multiple events that could cause the same state of digital data, there is an actual, true event that caused it, and one or more other events that did not.''
This means that rather than arriving at a single definitive inferred event $E'$, we may generate $k$ alternative events, denoted as $E'_j$ where $1 \leq j \leq k$. Each $E'_j$ represents a distinct interpretation of the evidence, each of which could potentially explain the observed data. These multiple instances of $E'$ highlight the complexity and ambiguity, where different sequences of events could produce similar artifacts.
The process involves not only constructing these alternatives but also systematically and repeatedly testing and eliminating hypotheses to converge on the most likely scenario while acknowledging that multiple interpretations may still be viable based on the available evidence. 
To test and eliminate hypotheses, \citet{casey2020standardization}'s `Strength of evidence scale' (C-Scale) may be used, and it may involve research into artifact interpretation and experiments to determine if a set of actions could produce the observed system changes.

\begin{table*}[t!]
\centering
\tiny 
\caption{Summary of Systematization of Knowledge (SoK) for Timeline-based Event Reconstruction (TER)}
\label{tbl:sok}
\begin{tabular}{|p{3.3cm}|p{3.4cm}|p{2.6cm}|p{0.17cm}|p{0.17cm}|p{0.17cm}|p{0.17cm}|p{0.1cm}|p{0.1cm}|p{0.1cm}|p{0.1cm}|p{0.1cm}|p{0.1cm}|p{0.1cm}|p{0.1cm}|p{0.1cm}|p{0.1cm}|}
\hline %
\multirow{2}{*}{\textbf{Paper}}                            & \multirow{2}{*}{\textbf{Focus area}}                  & \multirow{2}{*}{\textbf{Contribution type/Challenge}}            & \multicolumn{4}{c|}{\textbf{TER quadrant}} & \multicolumn{10}{c|}{\textbf{Data source category}} \\
\cline{4-17}
 &                                              &                                               & \textbf{Q1}    & \textbf{Q2}    & \textbf{Q3}    & \textbf{Q4}   & \rot{\textbf{Physical}} & \rot{\textbf{File system}} & \rot{\textbf{Multi sources}} & \rot{\textbf{Logs}} & \rot{\textbf{Other}}  & \rot{\textbf{Timestamp}} & \rot{\textbf{Analysis}} & \rot{\textbf{Mobile/IoT}} & \rot{\textbf{Volatile}} & \rot{\textbf{Network}} \\ \hline
\textbf{Sec. \ref{sec:event_reconstruction} Event reconstruction}                           &                                              &                                               &       &       &       &      &          &            &           &      &       &            &                        &            &          &         \\
\citet{lee2001henry}                                 & Foundational event reconstruction            & Conceptual framework                          & \checkmark     &       &       & \checkmark    & \smallnewmoon        &            &           &      &       &            &                        &            &          &         \\
\citet{carrier2004defining,carrier2004event}                     & Event-based investigation process            & Process model                                 & \checkmark     & \checkmark     &       &      &          & \smallnewmoon          &           & \smallnewmoon     &       &            &                        &            &  \smallnewmoon        & \smallnewmoon        \\
\citet{casey2011digital}                                      & Temporal, relational analysis                & Analytical framework                          &       &       & \checkmark     & \checkmark    &          &            & \smallnewmoon         &      &       &            &                        &            &          &         \\
\citet{chabot2015event}                             & Terminology, data volume                     & State-of-the-art review                       & \checkmark     &       &       & \checkmark    &          &            & \smallnewmoon         &      &       &            &                        &            &          &         \\
\citet{adderley2020interactive}                       & Temporal sequencing                          & Timeline correlation                          &       &       & \checkmark     &      &          &            &           & \smallnewmoon    &       & \smallnewmoon           &  \smallnewmoon                      & \smallfullmoon           & \smallfullmoon         & \smallfullmoon        \\
\citet{willassen2008hypothesis,willassen2008finding}                               & Hypothesis testing                           & Model-based reconstruction                    &       &       &       & \checkmark    &          &            &           & \smallnewmoon    &       & \smallnewmoon           &                        &            &          & \smallfullmoon        \\
\citet{batten2012hypothesis}                              & Hypothesis development                       & Reasoning methodology                         &       &       &       & \checkmark    &          & \smallfullmoon           &           & \smallnewmoon    &       & \smallfullmoon           &                        &            &          &         \\

\citet{xu2022visualizing}                                   & Knowledge graph reasoning                    & Visualization and reasoning model             &       &       &       & \checkmark    &          &            & \smallnewmoon         &      &       &            &                        &            &          &         \\ \hline
\textbf{Sec. \ref{sec:terms-def} Terminology}                                    &                                              &                                               &       &       &       &      &          &            &           &      &       &            &                        &            &          &         \\
\citet{neale2023fool}                                      & Artifact terminology harmonization           & Systematic terminology review                 &       & \checkmark     &       &      &          &            & \smallnewmoon         &      &       &            &                        &            &          &         \\
\citet{carrier2004defining,carrier2004event}                     & Event-based investigation process            & Process model                                 & \checkmark     & \checkmark     &       &      &          & \smallnewmoon          &           & \smallnewmoon     &       &            &                        &            &  \smallnewmoon        & \smallnewmoon        \\
\citet{hargreaves2012automated}                    & Event granularity                            & Event granularity                             &       & \checkmark     & \checkmark     &      &          & \smallnewmoon          &           & \smallnewmoon     &       & \smallnewmoon           &                        & \smallfullmoon           &          & \smallfullmoon        \\
\citet{marrington2007event}                          & Computer activity                         & Activity terminology                       &       & \checkmark     & \checkmark     &      &          & \smallnewmoon           &           & \smallnewmoon    &       & \smallnewmoon           &                        &            &          &         \\
\citet{neasbitt2014clickminer}                            & User interaction terminology                 & Interaction terminology                       &       &       & \checkmark     &      &          &            &           & \smallfullmoon     &      &            &                        &            &          & \smallnewmoon        \\
\citet{chabot2014complete}                              & Duration-based event definition              & Terminology refinement                        & \checkmark     &       &       & \checkmark    &          &            & \smallnewmoon         &      &       &            &                        &            &          &         \\
\citet{jaquet2021formalized}                  & Forensic event structure                     & Forensic event model                          & \checkmark     &       &       & \checkmark    &          &            & \smallnewmoon         &      &       &            &                        &            &          &         \\
\citet{harichandran2016cufa}                        & Artifact properties analysis                 & Artifact comparison                           &       & \checkmark     &       &      &          &            & \smallnewmoon         &      &       &            &                        &            &          &         \\
\citet{horsman2019raiders}                                    & Artifact as digital object              & Practical definition                          &       & \checkmark     &       &      &          & \smallnewmoon          &           & \smallnewmoon     &       &            &                        &            &          &         \\
\citet{casey2022crowdsourcing}                               & Artifact definition                      & Artifact catalog                            &       & \checkmark     &       &      &          &            & \smallnewmoon         &      &       &            &                        &            &          &         \\
\citet{lyle2022digital}                                & Artifact identification                          & Digital investigation techniques                          &       & \checkmark     &       &      &          &            & \smallnewmoon         &      &       &            &                        &            &          &         \\ \hline
\textbf{Sec. \ref{sec:model} Model for event reconstruction}                 &                                              &                                               &       &       &       &      &          &            &           &      &       &            &                        &            &          &         \\
\citet{ribaux2014police,ribaux2023police}                               & Forensic trace model                         & Trace-based model                             &       & \checkmark     &       &      & \smallnewmoon        &            &           &      &       &            &                        &            &          &         \\
\citet{vanini2023presentation}                              & Event source reliability                     & Reliability modeling                          & \checkmark     &       &       & \checkmark    &          &            & \smallnewmoon         &      &       &            &                        &            &          &         \\
\citet{vanini2024clock}                             & Time anchor model                            & Timestamp interpretation framework            & \checkmark     &       &       & \checkmark    &          &            &           & \smallnewmoon      &       & \smallnewmoon          &                        &            &          & \smallfullmoon        \\
\citet{carrier2006hypothesis}                                    & Investigation process model                  & Hypothesis-based model                        &       &       &       & \checkmark    &          & \smallnewmoon          &           & \smallfullmoon     &       &            &                        &            &          &         \\
\citet{hargreaves2009assessing}                                 & Evidence reliability testing                           & Reliability criteria                   &       &       &       & \checkmark    &          &            & \smallnewmoon         &      &       &            &                        &            &          &         \\
\citet{jaquet2021formalized}                  & Event structure                              & Formal event model                            & \checkmark     & \checkmark     &       &      &          &            & \smallnewmoon         &      &       &            &                        &            &          &         \\
\citet{hargreaves2024abstract}                         & Tool transparency                            & Tool capability model                         &       & \checkmark     & \checkmark     &      &          &            & \smallnewmoon          &      &       &           & \smallnewmoon                       &            &          &         \\
\citet{raghavan2013unitime}                          & Timestamp interpretation                     & Timestamp model                               & \checkmark     &       & \checkmark     &      &          &            &           & \smallnewmoon    &       & \smallnewmoon          &                        & \smallfullmoon           &           & \smallnewmoon        \\
\citet{hargreaves2012automated}                    & Timeline generation model                    & Timeline generation model                     &       & \checkmark     & \checkmark     &      &          & \smallnewmoon          &           & \smallnewmoon     &       & \smallnewmoon           &                        & \smallfullmoon           &          & \smallfullmoon        \\
\citet{studiawan2020automatic,studiawan2023event}                          & Event abstraction                            & Event abstraction model                       &       &       & \checkmark     & \checkmark    &          &            &           & \smallnewmoon     &       & \smallnewmoon           & \smallnewmoon                      &  \smallfullmoon          &          & \smallfullmoon        \\ 
\citet{carrier2004defining,carrier2004event}                     & Hypothesis-based investigation               & Hypothesis model                              & \checkmark     &       &       & \checkmark    &          & \smallnewmoon          &           &      &       &            &                        &            &          &         \\
\citet{gladyshev2004finite}                         & Event inference                              & FSM reconstruction                            &       &       &       & \checkmark    &          &            &           & \smallnewmoon    &       & \smallfullmoon           &                        &            &          & \smallfullmoon        \\
\citet{amato2017correlation}                               & Semantic evidence correlation                & Ontology-based model                          &       & \checkmark     & \checkmark     &      &          &            &           & \smallnewmoon    &       &            &                        &            &          &         \\
\citet{xu2022visualizing}                                   & Knowledge graph presentation                 & Reasoning model                               &       &       & \checkmark     & \checkmark    &          &            & \smallnewmoon         &      &       &            &                        &            &          &         \\ \hline
\multicolumn{2}{|l|}{\textbf{Sec. \ref{sec:challenges-in-reconstruction} Challenges stemming from environmental and process-related factors}}                  &                                              &                                               &       &       &       &      &          &            &           &      &       &            &                        &            &                   \\
\textbf{Sec. \ref{sec:incorrect_sys_time} Incorrect environment time}                  &                                              &                                               &       &       &       &      &          &            &           &      &       &            &                        &            &          &         \\
\citet{stevens2004unification}                                    & Misconfigured system clocks                  & Clock drift challenge                         & \checkmark     &       &       &      &          &            &           &      &       & \smallnewmoon          &                        &            &          &         \\
\citet{raghavan2013unitime}                          & Timestamp normalization and storage issues   & Timestamp interpretation framework            & \checkmark     &       &       &      &          &            &           & \smallnewmoon      &       & \smallnewmoon          &  \smallnewmoon                      & \smallfullmoon           & \smallnewmoon         & \smallnewmoon        \\
\citet{vanini2024clock}                             & Time anchor abstraction model                & Time anchor modeling                          & \checkmark     &       &       & \checkmark    &          &            &           & \smallnewmoon      &       & \smallnewmoon          &                        &            &          & \smallfullmoon        \\
\citet{kaart2014android}                           & Incorrect timezone data handling             & Time zone configuration             & \checkmark     &       &       &      &          &            &           & \smallnewmoon     &       & \smallnewmoon          & \smallnewmoon            & \smallnewmoon            &  \smallnewmoon        & \smallnewmoon        \\
\citet{schatz2006correlation,buchholz2007brief} & Network-induced skew, unsync clocks  & Distributed system time consistency & \checkmark     &       &       &      &          &            &           &  \smallnewmoon    &       & \smallnewmoon          & \smallnewmoon      & \smallfullmoon           &          &  \smallnewmoon       \\
\citet{henderson2009categorization}                                  & Clock skew in shared environments            & Network delay and skew                        & \checkmark     &       &       &      &          &            &           & \smallnewmoon     &       & \smallnewmoon          &                        &  \smallfullmoon &          & \smallnewmoon        \\
\multicolumn{2}{|l|}{\textbf{Sec. \ref{sec:configurations-and-implementations} Configurations and implementations}} &                                               &       &       &       &      &          &            &           &      &       &            &                        &            &          &         \\
\citet{adedayo2015ideal}                          & Log suppression, redirection                 & Log misconfiguration                          & \checkmark     &       &       &      &          &            & \smallnewmoon          & \smallnewmoon     &      &  \smallfullmoon          &                        &            &          &         \\
\citet{fernandez2022digital}                   & Absence of traceability in apps              & Limited logging capability                    & \checkmark     &       &       &      &          &            &           & \smallfullmoon     &       &  \smallfullmoon          &                        & \smallnewmoon          & \smallnewmoon         &         \\
\textbf{Sec. \ref{sec:environmental-anomalies} Environmental anomalies}                     &                                              &                                               &       &       &       &      &          &            &           &      &       &            &                        &            &          &         \\
\citet{studiawan2019survey}      & Unrecoverable system restarts                & Environmental disruption                      & \checkmark     &       &       &      &          &            &           & \smallnewmoon    &       &  \smallnewmoon          & \smallnewmoon                       & \smallfullmoon           & \smallfullmoon         & \smallfullmoon        \\
\citet{oh2022forensic}                                  & Sudden device restarts                       & Restart-induced log gaps                      & \checkmark     &       &       &      &          & \smallnewmoon           &           & \smallnewmoon    &       &   \smallnewmoon         &                        & \smallfullmoon           &          &         \\
\citet{marrington2011cat}                          & Program faults, data corruption              & Software instability                          & \checkmark     &       &       &      &          &  \smallnewmoon          &           & \smallnewmoon    &       &  \smallnewmoon          &  \smallnewmoon                      &            &          &         \\
\textbf{Sec. \ref{sec:data-fluctuation} Data fluctuation}                            &                                              &                                               &       &       &       &      &          &            &           &      &       &            &                        &            &          &         \\
\citet{sandvik2021towards}                             & Short lifespan of traces                     & Volatile trace loss                           & \checkmark     &       &       &      &          &            &           &      &       &            &                        &            & \smallnewmoon        &         \\
\citet{marangos2016time}                            & Evidence affected by operational cycles      & Temporal instability                          & \checkmark     &       &       &      &          &            &           &      &       &            &                        &            & \smallnewmoon        &         \\
\textbf{Sec. \ref{sec:post-event-period} Post-event period}                             &                                              &                                               &       &       &       &      &          &            &           &      &       &            &                        &            &          &         \\
\citet{gruber2023contamination}                              & Evidence altered during acquisition          & Contamination challenge                       &       & \checkmark     &       &      &          & \smallnewmoon           &           & \smallnewmoon     &       & \smallnewmoon           &                        & \smallnewmoon           & \smallnewmoon        &         \\
\citet{jaquet2021formalized}                  & Evidence fragility and impermanence          & Temporal evidence integrity         &       & \checkmark     &       &      &          & \smallnewmoon          &           &      &       &            &                        &            &          &         \\
\citet{khan2007framework}                                & Overwriting of data, log aging               & Aging challenge                               &       & \checkmark     &       &      &          & \smallnewmoon           &           & \smallnewmoon    &       & \smallnewmoon           &                        &            & \smallfullmoon         &         \\
\citet{soltani2019event}; \citet{schuster2007introducing}            & Metadata decay, inaccuracy                   & Artifact degradation                          &       & \checkmark     &       &      &          & \smallnewmoon          & \smallnewmoon          & \smallnewmoon     &       & \smallnewmoon           &                        &            & \smallfullmoon         & \smallfullmoon        \\
\textbf{Sec. \ref{sec:timeline-generation-tool-usage} Timeline}            &                                              &                                               &       &       &       &      &          &            &           &      &       &            &                        &            &          &         \\
\citet{patterson2012potential}                    & Cross-source correlation                     & Source integration challenge                  &       &       & \checkmark     &      &          & \smallnewmoon & \smallnewmoon          & \smallnewmoon    &       & \smallnewmoon           &                        & \smallnewmoon           &          &         \\
\citet{mohammed2016automated}                            & Data format diversity                        & Data normalization challenge                  &       &       & \checkmark     &      &      &  \smallnewmoon          & \smallnewmoon             & \smallnewmoon        &       &  \smallnewmoon          &  \smallnewmoon                      & \smallnewmoon          &  \smallnewmoon        &  \smallnewmoon       \\
\citet{horsman2019raiders}                                    & Artifact parsing complexity                  & Parser dependency challenge                   &       &       & \checkmark     &      &          & \smallnewmoon          &           & \smallnewmoon     &       &            &                        &            &          &         \\
\citet{soltani2017survey}                            & Missing / incomplete timestamps              & Extraction incompleteness                     &       &       & \checkmark     &      &          &            & \smallnewmoon         &      &       &            &                        &            &          &         \\
\citet{gomez2005using,levett2010towards}         & Correlation of heterogeneous data            & Multi-source correlation                      &       &       & \checkmark     &      &          &            & \smallnewmoon         &      &       &            &                        &            &          &         \\
\citet{kalber2013forensic,hargreaves2024abstract} & Tool transparency and automation limitations & Human-tool balance challenge                  &       &       & \checkmark     & \checkmark    &          &            & \smallnewmoon         &      &       &            &                        &            &          &         \\
\citet{bhat2021can}                                & Misconfigured analysis environments          & Tool setup challenge                          &       &       & \checkmark     & \checkmark    &          &            &  \smallnewmoon         &      &       &            & \smallnewmoon                      &            &          &         \\
\textbf{Sec. \ref{sec:q4:decisionmaking} Decision making}                               &                                              &                                               &       &       &       &      &          &            &           &      &       &            &                        &            &          &         \\
\citet{chabot2015event} & Data volume for timeline analysis & Scalability and overload challenge & &       &       & \checkmark     &          &            & \smallnewmoon          & \smallnewmoon     &       &            &                        &            &          &         \\
\citet{quick2014impacts} & Computational resource limitations & Resource requirement challenge & &       &       & \checkmark     &          &            & \smallnewmoon          & \smallnewmoon     &       &            &                        &            &          &         \\
\citet{buchholz2005design} & Event aggregation & Event abstraction for analysis & &       &       & \checkmark     &          &            & \smallnewmoon          &  \smallnewmoon    &       &            &                        &            &          &         \\
\citet{kiernan2009eventsummarizer} & Event summarization & Abstraction and streamlining & &       &       & \checkmark     &          &            &          &  \smallnewmoon     &       &            &                        &            &          &         \\
\citet{osborne2009enhancing} & Visualization accuracy & Visual representation integrity & &       &       & \checkmark     &          &            &           &      &       &            &  \smallnewmoon                      &            &          &         \\
\hline
\multicolumn{2}{|l|}{\textbf{Sec. \ref{sec:challenges_tampering} Challenges stemming from deliberate interference}}    &       &       &      &     		&          	&            &           &      	&      		&            				&             &            &  &   &      \\
\citet{casey2020standardization}                                      & Strength and scale of inference              & Evaluative opinion framework        &       &       &       & \checkmark    &          &            &           &      & \smallnewmoon     &            &                        &            &          &         \\
\citet{vanini2024clock} & Time manipulation, clock tampering					& Timeframe manipulation   	& \checkmark      &       &      &          &          &            &           & \smallnewmoon      &       & \smallnewmoon          &                        &            &          & \smallfullmoon        \\
\citet{mitre2023impair}	& Environment manipulation, disabled logging	& Environment tampering        & \checkmark      &       &      &     		&          	&            &           & \smallnewmoon     	&      		&            				&             &            &  &        \\
\citet{conlan2016anti}	& Erasure or alteration of evidence using tools 	&  Anti-forensics tool usage       & \checkmark      &       &      &     		&          	&            & \smallnewmoon          &      	&      		&            				&             &        &    &         \\	
\citet{palmbach2020artifacts}	& File and log manipulation using malware 		&  Malware-assisted anti-forensics      & \checkmark      &       &      &     		&          	&            & \smallnewmoon          &      	&      		&            				&             &            &      &   \\	
\citet{malhotra2015attacking}	& Service manipulation (e.g., NTP tampering)  	&  Service compromise       & \checkmark      &       &      &     		&          	&            &           &      	&      		&            				&             &            & &  \smallnewmoon        \\	
\citet{choi2021forensic}	& Post-event manipulation: logs, timestamps, files 	&  Artifact modification \& deletion       &       & \checkmark      &      &     		&          	&            &           & \smallnewmoon     	&      		&            				&             &            & &        \\		\hline
\multicolumn{17}{|l|}{Notes: \newmoon ~Mentioned in the paper \fullmoon ~Not specifically mentioned, but can be implemented using the data source} \\
\hline
\end{tabular}
\end{table*}

\section{Methodology for challenge identification} 
\label{sec:methodology-for-challenges}

To identify and categorize the challenges in event reconstruction, we followed a structured literature review process designed to balance breadth with relevance. The goal was not to exhaustively capture all existing work but to obtain a representative and insightful overview of the key challenges discussed in the field.

\begin{description}[noitemsep] 
    \item[Search strategy:] We defined a set of core search terms related to the topic: event reconstruction, timeline, timestamp analysis, digital forensics, correlation, challenges, and problems. These terms were combined using Boolean operators and phrasing variations (e.g., quotation marks for exact matches). Searches were conducted using Google Scholar, which indexes most major academic publishers (e.g., IEEE, ACM, Wiley, Springer) and relevant platforms such as DFRWS.org and arXiv.

    \item[Selection criteria:] For each query, we considered the first two pages of results (i.e., 20 entries). Articles were initially screened based on metadata displayed: title, author(s), publication venue, and two-line extract. If no direct reference to digital forensics was evident, the article was discarded. This filtering yielded a preliminary pool of approximately 200 articles.

    \item[Challenge extraction:] We extracted mentions of challenges primarily from the abstract and introduction sections, where such content is frequently summarized. Targeted keyword searches (e.g., challenge, problem, limitation) were also used within full texts to uncover implicit references.

    \item[Classification:] The identified challenges were then mapped onto a diagram, categorizing them according to the stage or context in which they occur within the event reconstruction process.
\end{description}

We also incorporated our domain expertise to address gaps in the literature, recognizing that some relevant challenges may not have been explicitly highlighted in existing works.

\paragraph{Limitations} The article collection and analysis were conducted manually, which may have led to the omission or misclassification of relevant articles. By restricting searches to Google Scholar and considering only the first two pages of results, important sources further down the list or from other databases may have been excluded. The focus on abstracts and introductions might have caused us to overlook challenges discussed deeper within the papers. Moreover, the subjective nature of challenge classification introduces potential bias based on the researchers' interpretations. Finally, the absence of automated or statistical tools for extraction and categorization limits the objectivity and comprehensiveness of the analysis. Despite these limitations, we believe the following sections offer a comprehensive and nuanced overview of the challenges.

\section{Challenges stemming from environmental and process-related factors}
\label{sec:challenges-in-reconstruction}

This section focuses on \emph{unintentional} challenges and the structure follows the diagram's flow, discussing each quadrant.

Note, that while we have strived to define the challenge categories as distinctly as possible, some overlap is inevitable due to the interconnected nature of these activities. Certain actions may reasonably fall into multiple categories, depending on the context. The categorization is designed to provide guidance rather than enforce strict mutual exclusivity.
\black

\subsection{Q1: Timeframe of interest}
Four areas have been identified:

\subsubsection{Incorrect environment time}
\label{sec:incorrect_sys_time}
Clock-related challenges originate from the system time which is used to derive timestamps. If the clock is incorrect, all timestamps originating from this clock are incorrect  \citep{stevens2004unification,raghavan2013unitime, vanini2024clock}. 

\begin{description}[noitemsep]
    \item[Clock skew:] Skew refers to the difference in time readings between different systems. One reason for clock skew could be propagation delays which may occur due to network delays \citep{schatz2006correlation,henderson2009categorization} or due to synchronization problems, e.g., NTP servers providing incorrect times \citep{buchholz2007brief,hampton2016timestamp}.
    
    \item[Clock drift:] Drift is the gradual deviation of a clock from the correct time, often caused by factors such as changes in temperature, voltage fluctuations, or inherent defects in the clock circuitry \citep{sandvik2018reliability}. 
    Clock drift may exacerbate over time. As drift accumulates, the discrepancies between different systems' clocks can grow, making it increasingly difficult to correlate events across environments \citep{becker2008implications}.
    
    \item[Time zone changes:] As systems traverse different time zones, whether due to travel or daylight-saving time changes, the system time may change \citep{stevens2004unification}. 
    This adjustment process can also be error-prone, e.g., due to an inaccurate time zone database \citep{kaart2014android}. 
    Compared to skew and drift, the range is significantly larger, i.e., hours instead of seconds. Typically this is only relevant where local time is stored in a data structure rather than storing UTC.

\end{description}

Note that virtual environments come with their challenges which are beyond the scope of this article but have been discussed in \citet{vmware2008timekeeping}. 
\black

\subsubsection{Configurations and implementations}
\label{sec:configurations-and-implementations}

Environments, systems, and application configurations define how/what data is generated, stored, and logged. These configurations comprise a wide range of settings, including logging levels, storage policies, network settings, and security controls. 

\begin{description}[noitemsep]
    \item[Suppression/deletion:] Conservative default settings can result in insufficient logging, leading to missing artifacts, e.g., database logs prioritizing space efficiency over detail \citep{adedayo2015ideal}. Systems may also be configured to suppress artifacts, such as private browsing \citep{fernandez2022digital}, or delete them, such as printer jobs removed after completion \citep{gladyshev2004finite} or when an application is closed.
    
    \item[Inconsistent implementations:] Different resolutions lead to inconsistencies, e.g., timestamps recorded in hh:mm vs.~hh:mm:ss format \citep{song2016cleaning}. File systems, drivers, and implementations may behave differently leading to unpredictable behavior \citep{bang2009analysis,nordvik2022time}.

\end{description}

\subsubsection{Environmental anomalies}
\label{sec:environmental-anomalies}

Environments may not behave as expected leading to the destructing of evidence or the not-creation of artifacts:

\begin{description}[noitemsep]
    \item[(OS) Crashes:] A crash (system, application) can result in the loss or corruption of artifacts, potentially leaving logs incomplete and missing key events \citep{studiawan2019survey, oh2022forensic}. Detecting crashes can be challenging, particularly if the logging mechanisms themselves are compromised during the crash. Crashes may also lead to restart anomalies such as services or applications that are supposed to start automatically failing to do so potentially altering the way subsequent events are logged.

    \item[Software bugs:] Bugs in software may cause errors in data logging, such as incorrect timestamps or missing events \citep{marrington2011cat}. 

    \item[Resource exhaustion and failure:] Environments under heavy load may fail to log events properly due to resource constraints, leading to delayed or missed entries in the event data. Failures, including hardware malfunctions, can lead to inadequate data \citep{marrington2011cat}. 
\end{description}

\subsubsection{Data fluctuation}
\label{sec:data-fluctuation}

Data may not be accessible due to or only with additional burden:

\begin{description}[noitemsep]
    \item[Data volatility:] Volatile data, such as RAM content or network traffic, is lost if the $\Delta$ is too large. In addition, IoT devices often have resource constraints resulting in short-lived data \citep{sandvik2021towards}. 
    In cloud environments, VMs can be easily deleted including their logs \citep{marangos2016time}.

    \item[Environment bounds:] The changes resulting from an event may be distributed across multiple locations, including cloud environments, resulting in fragmented evidence that is challenging to collect and analyze \citep{nist2014nist, joseph2019digital,manral2019systematic}. 

\end{description}

Even with the cooperation of external service providers, data cannot be recovered, particularly when logging is explicitly disabled, as is often the case with many VPN services.

\subsection{Q2: Post-Event Period}
\label{sec:post-event-period}

This period relates to the influence of time on the changes left behind after an event.

\subsubsection{Subsequent events impacting changes}
Over time the changes generated by the primary event are altered by subsequent events (referred to as intrinsic events by \citet{jaquet2021formalized}, or evidence dynamics by \citet{gruber2023contamination}).

\begin{description}[noitemsep]
    \item[Deletion:] Initial changes may disappear due to subsequent events. Examples are rotating logs \citep{sandvik2021towards}, temporary files, routine cleanup tasks, or reboots.

    \item[Alteration/overwriting:] Subsequent events can modify or replace existing data. For instance, \citet{khan2007framework} mention that much of the application footprint is rewritten each time the application runs. Routine file operations, such as automatic backups or updates, may also overwrite metadata, configurations, or timestamps \citep{soltani2019event}. 
    
\end{description}

\subsubsection{Aging and degradation}
Digital artifacts and physical devices are susceptible to degradation, affecting their reliability and accessibility. This degradation can manifest as file corruption, obsolescence of file types, or the deterioration of storage media.
Furthermore, changes in software, file formats, or logging systems can introduce additional challenges. As schemas evolve, inconsistencies in log formats may emerge, complicating the process of reconciling older and newer data entries. Backward compatibility issues also arise when outdated systems or logs are incompatible with modern tools, requiring extra effort to ensure that historical data remains interpretable and consistent across different versions \citep{schuster2007introducing}.

\subsection{Q3: Timeline}
\label{sec:timeline-generation-tool-usage}

This third quadrant summarizes all timeline-related challenges. We decided to include the trans-boundary boxes, i.e., timeline generation (Q2-Q3) and timeline analysis (Q3-Q4), in this section as we think they are closer related to the timeline.

\subsubsection{Timeline generation}
\label{sec:timeline_generation}
Data comes from various systems, including traditional computing environments and a growing number of IoT devices, each with distinct structures, conventions, and formats \citep{patterson2012potential, mohammed2016automated}. 
This increasing \emph{heterogeneity} of both data sources and devices causes several challenges.

\begin{description}[noitemsep]
    \item[Artifact/timestamp extraction:] Extracting data presents an ongoing challenge, as tools must be continuously updated to accommodate new and evolving software \citep{horsman2019raiders}. The acquisition process can introduce alterations, particularly when conducted on live systems, such as during memory dumps \citep{soltani2017survey, gruber2023contamination}. 

    \item[Normalization:] This involves converting diverse data types, such as logs, databases, and sensor outputs, into a standardized structure that enables comprehensive analysis \citep{han20205w1h}. This can be challenging due to different timestamp formats, timestamp resolutions, and timezone settings. Timestamp formats can also change over time, meaning timestamp normalization needs to be updated over time and handle older and newer formats. 
    
    \item[Contamination and process problems:] Evidence might be unintentionally modified during collection or handling, e.g., failing to use a write blocker \citep{gruber2023contamination} or corrupt software, leading to data contamination. Similarly, lapses in maintaining a proper chain of custody can result in evidence being mishandled, misplaced, or questioned in terms of authenticity and reliability. 

    \item[Source combination:] 
    Combining data from multiple sources to create a unified perception is challenging, especially when sources have different levels of reliability or granularity \citep{gomez2005using,levett2010towards}. 
\end{description}

\subsubsection{Tool capabilities and usage}
Balancing automated tools with manual analysis is essential yet challenging. While automation expedites the process, it may overlook nuances that a human analyst would catch \citep{kalber2013forensic} and can introduce various types of error \citep{hargreaves2024abstract}. 

\begin{description}[noitemsep]
    \item[Usage challenges:] Incorrect settings or carelessness can lead to incorrect results. For example, errors in the configuration of the tools have been shown to result in inaccurate extractions of digital evidence, which can impact the credibility of the findings \citep{bhat2021can}. The transition to a new tool may lead to misinterpretation as tools may interpret/visualize data differently. Some features of tools also do not help in reducing chances of investigator misinterpretation (see \cite{hargreaves2024abstract}), e.g., if a tool provides an automated result of a Google search occurring, this is easy to interpret the event occurring as a fact rather than Google search data being present. This is an event reconstruction process, with all the uncertainty that could be present, as discussed in Sec.~\ref{sec:q4:decisionmaking}. 
    Tools can conflate facts with interpretation within their interfaces. 

    \item[Transparency:] Many tools operate as black-boxes making it unclear how artifacts are handled. Transparency of functionality is critical, as proprietary processes can influence assumptions or conclusions, leading to misinterpretation.
    
    \item[Handling volume:] Tools may have limits on the amount of data they can process or the complexity of queries, leading to unnoticed gaps in analysis, e.g., a tool limited to analyzing  5,000 files at once. 
    Consequently, validation is essential, but challenging, given the rapid change of artifacts \citep{horsman2018couldn,arshad2018digital}. 

    \item[AI-powered examination:] AI-powered tools introduce complexities regarding explainability and transparency, not just of the models but of training data. Recent approaches such as LLMs are also problematic due to their non-deterministic nature and in many cases opaque training data and processes. These tools can produce inaccurate or misleading outputs, such as AI-generated errors or `hallucinations' which can affect the analysis \citep{scanlon2023chatgpt}.
\end{description}

Developers aiming to create tools should consider the seven criteria outlined by \citet{chabot2015ontology}, which provide a comprehensive framework for ensuring an efficient reconstruction tool.

\subsection{Q4: Decision Making} 
\label{sec:q4:decisionmaking}

Q4 involves the generation and testing of hypotheses based on the timeline. This is critical and \cite{hargreaves2009assessing} goes as far as defining a digital investigation as ``a process that formulates and tests hypothesis using digital evidence'' with the prior stages facilitating this goal. Some areas of this are explored, e.g., timeline analysis, but others, such as hypothesis forming and testing are less frequently discussed.

\subsubsection{Timeline analysis}
Although the processing is mostly done using tools, this section highlights challenges originating from the processing of timeline entries. 

\begin{description}[noitemsep]
    \item[Volume of data:] 
    The extensive amount of information (number of entries in the timeline) makes the analysis time-consuming \citep{chabot2015event} and overloads examiners. 
    Additionally, significant resources are needed to extract, process, and store this data, including computational power, storage capacity, and advanced data management tools \citep{quick2014impacts}.

    \item[Aggregation, organization and visualization:] 
    Techniques such as combining related events into cohesive units (sometimes called high-level events or super events) \citep{buchholz2005design, kiernan2009eventsummarizer, hargreaves2012automated, inglot2014enhanced, raju2017closer} can streamline analysis but may result in the loss of granularity or context. Similarly, visualizations \citep{osborne2009enhancing} require consideration to ensure that they accurately represent the data without oversimplifying or distorting the information. The volume of the raw data can be a challenge to visualize and reduction of the data before visualization is meaningful may be necessary, e.g., \cite{hargreaves2012automated}.

    \item[Correlation:] The process of establishing meaningful relationships between disparate timelines entries is fraught with difficulties, especially when data originates from various sources or formats \citep{schatz2006correlation} or times across environments are not synchronized \citep{marangos2016time}. Detecting and validating these connections requires experience and meticulous attention \citep{amato2017correlation}.  For example, incorrect handling of local time vs.~UTC can disrupt the sequencing of events, particularly in global systems where data spans multiple time zones \citep{buchholz2007brief}. 
    Verifying data across different sources and formats is challenging but necessary to ensure the accuracy and completeness of the reconstructed timeline.

\end{description}

\subsubsection{Interpretation, trust and integrity}
Ensuring that data is accurate and trustworthy is fundamental \citep{neale2022case}. Determining which sources to trust and how to weigh them can significantly affect the reliability of the reconstruction. This challenge becomes even more pronounced when different sources report the same event but provide inconsistent or conflicting details, leading to uncertainty. 

\begin{description}[noitemsep] 
    \item[Interpretation:] Investigators work with a static set of data which includes evidence and irrelevant information generated by subsequent activities or during investigative processes \citep{roux2022sydney}. Misinterpretation can arise from factors such as incorrect ordering, aggregation, or filtering of entries, leading to distortions in the reconstructed narrative but also from unawareness of an examiner, i.e., insufficient knowledge of an event or timestamp \citep{boyd2004time}. 

    \item[Untrusted internal sources:] The presence of anti-forensic tools \citep{conlan2016anti} or tampering indicators, such as manipulated timestamps or hidden data, raises suspicion about the authenticity of the evidence\footnote{  We decided to include this challenge here and not in Sec.~\ref{sec:challenges_tampering} (deliberate interference) as the presence of these tools does not necessarily mean that they were executed.}. According to \citet{neale2023fool}, detecting and addressing such tampering is crucial to maintaining trust in the evidence (more in Sec.~\ref{sec:challenges_tampering}).

    \item[Untrusted external sources:] Combining data from external sources, such as cloud services, introduces additional challenges. When the integrity of these sources cannot be independently verified, especially due to possible alterations in transit or at rest, the reliability of the event reconstruction may be compromised \citep{battistoni2016cure}. 
\end{description}

\subsubsection{Knowledge and perception bias}
Investigators may interpret evidence differently based on their prior knowledge, experience, or expectations, which can lead to skewed interpretations of the data. Perception and decision bias may cause certain patterns or details to be overlooked. 

\begin{description}[noitemsep]
    \item[Artifact interpretation knowledge:] Previous knowledge may become outdated due to the release of a new operating system, or new version of an application \citep{horsman2019raiders}. 
    Examiners may be unaware of certain behaviors (e.g., \citep{thierry2022systematic} identified multiple unexpected and non-compliant behaviors of timestamps).
    Limitations in knowledge reduce the investigator's ability to generate viable alternative hypotheses that would produce the same artifacts. 

    \item[Algorithmic bias:] 
    Tools operate based on algorithms that might make certain assumptions or prioritize specific types of data, which can introduce biases into the reconstructed events \citep{jinad2024bias}. For instance, an AI-powered tool may be biased due to unbalanced training data. 

    \item[Human bias:] 
    Analysts may bring their own preconceptions into the analysis, influencing how they interpret and prioritize different events \citep{kang2013digital}. This can lead to confirmation bias, where analysts might favor hypotheses that align with their pre-existing beliefs or expectations, unintentionally skewing the analysis \citep{kassin2013forensic}. 
\end{description}

\subsubsection{Complexity in testing hypotheses}
Testing hypotheses against a timeline is complex, especially when considering all the aforementioned challenges. 

\begin{description}[noitemsep]

    \item[Multiple interpretations:] Evidence may be open to multiple interpretations, making it difficult to draw definitive conclusions and infer events from the past. This ambiguity can lead to varied interpretations of the same data, which impacts the ability to test hypotheses with certainty.
    Effective hypothesis testing must address temporal inaccuracies or manage the inherent uncertainty that arises from imperfect data such as log files \citep{latzo2019characterizing}. 

    \item[Defining error:] \citet{hargreaves2009assessing} discusses that error in event reconstruction can be defined as ``the difference between the inferred history and the true history of the examined digital evidence''. This error cannot necessarily be expressed as a definite value, e.g., $x \pm y$, but can be expressed as uncertainty (possible error) in the inferred events, i.e., alternative possible hypothesized events that explain the current state of the examined digital evidence. Communicating these uncertainties transparently is vital to ensure that conclusions drawn are appropriately qualified and reflect the limitations of the available evidence.

\end{description}

\section{Challenges stemming from deliberate interference}
\label{sec:challenges_tampering}
To complement the previous section, this one outlines challenges stemming from deliberate actions such as backdating, erasing, or wiping, to hide activities \citep{casey2020standardization}. While it may not always be the case, for this work we assume that the investigative body and tool vendors are free from insider threats. Therefore, challenges are limited to the \emph{reality}.

As already pointed out in Sec.~\ref{sec:challenges-in-reconstruction}, some overlap of challenges is inevitable due to the interconnected nature of these activities.
\black

\subsection{Q1: Timeframe of interest}
\label{sec:manipulated_environments}
Interference with the environment can be conducted before the event occurs, with the intent to complicate investigations. Such interference often seeks to generate misleading artifacts or prevent their creation altogether, e.g., examples under `defence evasion' in the MITRE ATT\&CK Matrix\footnote{\url{https://attack.mitre.org/tactics/TA0005/}}.

\begin{description}[noitemsep]
    \item[Time manipulation:] An adversary may turn off set time and date automatically and actively manipulates the system time or timezone \citep{vanini2024clock}. 
    Even when detected, distinguishing between accidental misconfigurations and deliberate tampering remains difficult.

    \item[Environment manipulation:] 
    It is possible to disable or tamper with logging mechanisms, preventing activities from being recorded. Similarly, security tools may be compromised or altered \citep{mitre2023impair}.
    Decoys such as fake accounts or planted traps such as cleanup scripts may be used to further obscure activities.

    \item[Anti-forensics and malware:] 
    Adversaries may use software to obscure their actions. For instance, anti-forensic tools erase or alter evidence \citep{conlan2016anti} or rootkits and malware to cover access and manipulations to files and logs \citep{palmbach2020artifacts}. Anonymization services such as VPNs and TOR hide the attacker’s origin, making it difficult to trace activities

    \item[Service manipulation:] 
    Instead of manipulating an environment directly, an adversary may compromise utilized services. For instance, by manipulating the NTP service, an attacker can change the system time \citep{malhotra2015attacking}. Another example would be a compromised update server. 
\end{description}

\subsection{Q2: Post-Event Period} 
Post-event one may \textbf{manipulate or delete metadata or content} such as altering timestamps, modifying log entries, or deleting critical files (e.g., remote wiping of mobile devices). Logs and other files are often not protected against alternation or deletion \citep{choi2021forensic}. 
Active tampering and manipulation of artifacts present some of the most challenging obstacles in event reconstruction and the risk of misinterpretation increases \citep{casey2020standardization} especially when performed from advanced persistent threads.

\section{Key findings}
\label{sec:keyfindings}

This section summarizes the key findings identified in the foundational sections \ref{sec:event_reconstruction} to \ref{sec:model}, and the challenge identification sections  \ref{sec:challenges-in-reconstruction} and \ref{sec:challenges_tampering}:

\begin{enumerate}[noitemsep]  

\item The terms ``event'' and ``artifact'' in digital forensics are defined inconsistently across existing studies and it leads to ambiguity in their usage.

\item Event reconstruction relies on modeling two critical intervals: the timeframe of interest ($T$) where events occur, and the post-event period ($\Delta$) where subsequent changes may overwrite or obscure evidence.

\item Event reconstruction is highly affected by unintentional challenges such as incorrect system time, insufficient logging, environmental anomalies, and data volatility.

\item Subsequent events can delete, overwrite, or degrade digital artifacts; so they reduce the availability and reliability of evidence over time.

\item Timeline generation faces challenges from data heterogeneity, software updates, extraction errors, normalization issues, and tool limitations. 

\item Event reconstruction requires careful hypothesis generation and testing, but faces challenges from data volume, correlation complexity, trust issues, and investigator bias.

\item Deliberate actions such as time manipulation, anti-forensics, and post-event tampering can alter or destroy digital evidence and make event reconstruction even more challenging.

\item Several research directions have emerged to address challenges in event reconstruction, including forensic readiness, improved artifact extraction, timeline verification, tamper detection, AI/NLP integration, and advanced analysis techniques.
\end{enumerate}

\section{Discussion and research gaps }
\label{sec:futureResearchDirections}

From the previous sections, the summary of key findings, and Table \ref{tbl:sok} (which provides a mapping of the focus areas in Sections \ref{sec:event_reconstruction} to \ref{sec:challenges_tampering}, against the quadrants in Figure \ref{fig:diagram}, illustrating the distribution of existing research) it is possible to infer general research gaps. However, this section highlights selected significant challenges and proposes specific potential avenues for future research.  

The section is organized by quadrant of the TER-model, demonstrating the utility of the model as an organizational tool.
Given the vast body of literature, it is not feasible to reference every relevant article. Therefore, we focus on studies from our initial collection as well as recent works.

One general point, is that throughout the TER-model (Q1-Q4) a broad research gap is the understanding and handling of uncertainty, from system configuration through to a reliance on examiner knowledge for hypothesis generation and testing. This is considered an ongoing limitation to the process that requires addressing. 

\begin{fancybox}
\small\textbf{Research Gap 1.} Uncertainty is potentially introduced throughout the model and research into handling it at each stage, and how it could propagate is needed.
\end{fancybox}

\subsection{Q1: Timeframe of interest}

Digital forensic readiness is a proactive approach ensuring systems and networks are prepared to efficiently collect, preserve, and analyze evidence when a security incident occurs \citep{sachowski2019implementing}. Forensic readiness for event logging has been researched, as demonstrated by \citet{reddy2013architecture} and \citet{kebande2018novel}. To support forensic readiness, administrators should activate extended logging, which records additional data and audit trails. Moreover, operating system developers could still provide more comprehensive system-related logs \citep{rivera2019towards} but this conflicts with privacy-centric approaches expected from consumers.

This also has anti-forensics implications. If an attacker deletes logs (one of the primary sources for event reconstruction), investigators must first recover them (as discussed in Q2/Q3). To address this, security measures such as centralized or encrypted log servers could be implemented in systems where this is feasible, and even advanced techniques such as blockchain can be used to mitigate anti-forensic techniques \citep{klos2020securing}.

\begin{fancybox}
\small\textbf{Research Gap 2.} Forensic readiness needs further development, and more creative solutions need researching to achieve similar goals on `unmanaged' systems where forensic readiness solutions cannot be deployed. 
\end{fancybox}

\subsection{Q2: Post-event period}

In evidence seizure, timing has an effect during forensic investigations. This affects if volatile artifacts are captured if not done on time,  e.g., credentials stored in memory. Secondly, challenges related to cloud environments imply any delays in data acquisition may effortlessly cause the loss of crucial evidence, e.g., \citet{alqahtany2016forensic} discuss evidence that supports the need for timely acquisition. 
There is also the issue of long-term log retention by internet service providers, which may be important in some cases \citep{khan2016cloud}. Mandating extended retention ensures information can be accessed after an incident, but conflicts with privacy regulations. There are also `awareness' concerns. For victim systems, communication is crucial to ensure device owners minimize interactions with devices containing potential evidence. The same applies to examiners, where changes to the evidence should be anticipated and minimized from a data preservation/acquisition perspective \citep{gruber2023contamination}.
Moreover, recent work by \citet{SPICHIGER2025301867} highlights that preservation should not be narrowly focused on the trace itself but must also consider the reference environment in which the trace was produced. As systems evolve, e.g., through software updates, operating systems, or third-party services, insufficient preservation of reference data can result in a loss of contextual meaning and increase the uncertainty of later reconstructions. Expanding the definition of preservation to include such reference data is therefore essential in environments where evidence may need to be interpreted long after the fact.

\begin{fancybox}
\small\textbf{Research Gap 3.} 
There is little work on the persistence of artifacts, and determining if the absence of data is due to configuration, tampering, or simply the passing of time. Work in this area could reduce this aspect of uncertainty within the model and process, and provide practical advice on the temporal boundaries of useful preservation periods. 
\end{fancybox}

\subsection{Q3: Timeline}
This aspect of event reconstruction has received the most attention and many articles and concepts have been discussed.

\begin{description}[noitemsep]

\item[Continuous updates/improvement to timestamp extraction:]
Files and formats containing timestamps are subject to change. Ongoing research that tracks these changes and uncovers new timestamp sources provides the foundational data necessary. This means ongoing `artifact research' (as defined by \cite{breitinger2024dfrws}) is critical.

\item[Integration of non-explicit timing information:]
\citet{dreier2024beyond} discussed implicit timing (e.g., ordering of log file entries) to detect inconsistencies in an automated way. A second possibility is digital stratigraphy, as defined by \citet{casey2018digital}, and further implemented in \cite{schneider2024applying}, which is a method that takes advantage of file systems and the behavior of their allocation algorithms. By analyzing the logical position of files on a disk, investigators can infer potential events, provided they understand how the file system allocates those files. This knowledge enables the reconstruction of hypothetical sequences of events based on file placement.
These are still early implementations, and additional work is needed to evaluate more variations in environments, file systems, drivers, and behavior patterns. 

\item[Timeline representation:]
Timelines are mostly flat, i.e., textual files in chronological order. The community should explore alternatives. 
For instance, an ontology-based approach improves event reconstruction by providing a structured and formal representation of data, which helps standardize and automate the analysis process \citep{bhandari2020ontology}. An ontology captures the semantic relationships between events, objects, and subjects, allowing investigators to infer new facts, identify correlations between events, and visualize data more effectively \citep{chabot2015ontology,turnbull2015automated}. We should also reconsider visualizing timelines, moving beyond the frequently used basic bar charts counting the number of events within defined timeframes, and exploring AR or VR.

\item[Automated timeline verification:] 
\citet{willassen2008timestamp} introduced a hypothesis-based approach where investigators create clock hypotheses to model historical clock values and test their consistency with timestamp evidence. \citet{vanini2024clock} suggested using time anchors (i.e., artifacts that include internal and external timestamps) and looking for anomalies. 
Research efforts need to continue to build verification methods that allow us to identify whether the timeline is out-of-sequence (irregularities found) or likely correct.

\item[Tamper detection:]
\citet{galhuber2021time} found that timestamp forgery tools may introduce detectable changes, such as reducing timestamp accuracy from nanoseconds to seconds. Among the tools they evaluated, only one was capable of modifying the full range of file system timestamps on Windows. \citet{andrade2020expose} noted that \$FN timestamps are typically modified only by the Windows kernel and are generally unaffected by anti-forensic timestomping tools, offering an example of a timestamp that is harder to manipulate during event reconstruction. \citet{jang2016understanding} presented a method to detected time manipulation in NTFS file system.
More general experiments as conducted by \citet{schneider2020tampering, schneider2022prudent, vanini2024strategies} show that the probability of detecting it is high, especially when it concerns file metadata. One reason is that it is difficult to forge a timestamp without causing subsequent inconsistencies.
While some progress has been made in detecting tampering, this area still requires further exploration and automation. Ideally, a tool should be capable of analyzing a timeline and automatically highlighting all potential tampering events.

\end{description}

\begin{fancybox}
\small\textbf{Research Gap 4.} Advances in timeline generation research are still needed in multiple areas: from artifact research, integration of non-timestamp-based timing information, visualization of timelines,  and detecting inconsistencies and tampering.
\end{fancybox}

\subsection{Q4: Analysis and investigative conclusions}
This includes the timeline analysis which bridges Q3 and Q4 since it may revisited as part of Q4 hypothesis testing. 

\begin{description}[noitemsep]

\item[Timeline analysis:]
Efforts focus on methods to reduce and manage data, including techniques for filtering, labeling, and aggregating data. Flagging entries that match certain criteria can be performed, or more complex approaches such as discussed by \citet{hargreaves2012automated,studiawan2020sentiment} where patterns of events are bundled to provide multiple entries that support an event reconstruction. This reduces large timelines to more manageable sets of interesting events, but as they are inherently a reduced set, switching back to the lower-level entry view is an important feature to retain to see inferred events in context and show provenance of the reconstructed event. 
A limitation discussed by \cite{hargreaves2012automated} is the need to manually create the patterns that need to be matched based on research and experience. Better centralized documentation of the expected changes from sets of actions in different environments, similar to \cite{casey2022crowdsourcing, grajeda2018experience} and integration into a standard timeline analysis tool would make timeline analysis more accessible. 

Visualization is also a vital additional layer of abstraction to help make sense of the large amounts of data, and can be a valuable tool to assist with analysis, e.g., to support timeline-based cross drive analysis \citep{patterson2012potential}.

An increased availability of ground truth data sets with annotation of the actions carried out would assist with developing analysis plugins for tools \citep{grajeda2017availability}. Automated event inference, either using machine learning, or through automation in digital forensic experimentation to carry out actions and record the resulting traces may help with this.

\item[Artifact reliability:]
If the timeline contains conflicting information i.e., at least two artifacts provide conflicting information, a resolution is needed. Automation in identifying accurate artifacts would be advantageous. One possibility is to compare artifacts and assess their reliability, e.g., the ease of manipulating an artifact \citep{vanini2024strategies}. 
\cite{hargreaves2012automated} began work on handling conflicting artifacts, where each inferred high-level event was assigned a series of expected artifacts. On a match, the supporting \emph{and contradictory} timeline entries were stored within the inferred event, highlighting entries that were expected but absent, forming the basis for the evaluation of reliability assessment. 
\cite{casey2011digital} discusses the number of independent sources and their resistance to tampering as part of the C-Scale, but if this were to be more strictly quantified, e.g., with Bayesian networks for example \citep{kwan2008reasoning}, in terms of assigning weight to expected artifacts, other factors may have an impact. For example `artifact longevity', i.e., how long an artifact is known to persist may allow appropriate weight to be given to the absence of specific, expected, hypothesis-supporting information. It remains unclear how appropriate precise numerical assessments in event reconstruction are.

\item[AI integration:]
The use of AI for digital forensics is becoming more common \citep{du2020sok,jarrett2021impact}. AI can help analyze and identify digital evidence \citep{henseler2023chatgpt,sreya2023enhancing} or aid investigators in writing forensic reports \citep{michelet2024chatgpt}. As discussed by \citet{scanlon2023chatgpt}, LLMs may help with event analysis, such as suspicious activities or attack identification. However, they may hallucinate when responding to investigator questions. Future work should focus on evaluating and validating this new technology for forensic purposes. 
Others have tried to apply AI techniques to accelerate the process, e.g., by searching for anomalies \citep{studiawan2017graph,studiawan2021anomaly} or relevant artifacts \citep{du2020automated,markova2022detection}.

\item[Natural Language Processing (NLP) integration:]
NLP may support timeline analysis as each event is represented by a descriptive message. These messages contain valuable information that can be extracted and analyzed. By applying traditional NLP techniques, such as sentiment analysis \citep{silalahi2023transformer-sentiment,studiawan2020sentiment}, named entity recognition \citep{silalahi2023dfler,silalahi2023transformer-ner,studiawan2023rule}, and information extraction, researchers can derive insights. For future research, there is potential to explore other NLP methods to enhance the field. For instance, topic modeling and dependency parsing could be employed to gain deeper insights into events and establish relationships between them.

\item[Process mining:]
Event reconstruction is a common task in process mining \citep{weijters2001process, jurgensen2021trace}, though it is typically applied to business process logs \citep{nguyen2019event}. However, the domain faces similar challenges. For example, \citet{dixit2018detection} describe a set of timestamp-based indicators for identifying event ordering imperfections in logs and present a method for resolving these issues using domain knowledge. Therefore, future research could explore various process mining techniques \citep{van2016data} for forensic event reconstruction.

\item[Training and education:]
Specialized training and continuous education play a key role in ensuring investigators can handle complex cases and maintain the admissibility of evidence in court \citep{jahankhani2014digital}. However, cognitive biases and human errors can impact the integrity of findings, but some techniques can be used to mitigate this, e.g., collaborative approaches, such as the 4-eye principle—where at least two individuals review the findings. More research is needed to explore how collaborative techniques and advanced decision-support systems, including AI-assisted tools, can further minimize human errors and biases, ensuring more reliable and transparent event reconstruction processes.
\end{description}

\begin{fancybox}
\small\textbf{Research Gap 5.}
The challenge of performing efficient and effective timeline analysis remains. Handling the volume of extracted timestamps in an effective way is needed (Q3/4), which could include technological solutions such as performance improvements or AI based filtering, but also process changes, where the `extract everything' model needs research to ensure it is still the most appropriate approach. 
\end{fancybox}

\begin{fancybox}
\small\textbf{Research Gap 6.}
    Automation is likely the only practical way to handle the challenge of inferring events at scale (Q4), but how to handle the practical research challenge of automated inference of events from timeline entries that are subject to operating system, application, and environmental changes earlier on in the process (Q1,Q2) is challenging. 
\end{fancybox}

\begin{fancybox}
\small\textbf{Research Gap 7.} Ensuring and communicating a clear delineation between extracted timestamp values as facts, and inferred events as working hypothesis, in both research and in forensic tooling (Q4), requires work from digital forensic scientists, and potentially UX experts to clearly communicate residual uncertainty.
\end{fancybox}

\section{Conclusions}
\label{sec:conclusions}
Event reconstruction is a critical part of the digital forensic process, yet the process and terminology are vague and inconsistent. 
This work has shown that this mixture of terms can be unified and as a result, a systematic organization of issues associated with timeline-based event reconstruction can be compiled. 
When an event reconstruction is completed, these potential issues can be considered and evaluated as to whether they may have influenced the result of the reconstruction. Aside from practical uses, it has also allowed clear future directions in event reconstruction research to be identified.

While some of these identified challenges will be obvious to seasoned investigators, there is a need within digital forensics, to formalize definitions and make explicit that which is currently tacit. This provides the foundation for more formal and potentially future quantitative evaluation of the trustworthiness or indeed reliability of reconstructed events in a digital forensic investigation.

\section*{Acknowledgments}
We acknowledge Eoghan Casey for the comments and feedback. The authors also thank C\'{e}line Vanini for the initial diagram and discussions.


\section*{Disclosure of AI-assisted writing tools} 
Some authors utilized ChatGPT-4 to assist in revising, condensing text, and correcting grammatical errors, typos, and awkward phrasing. All AI-generated suggestions were carefully reviewed and modified as necessary to ensure they aligned with the authors' intended meaning before being incorporated into this paper.

\section*{Declaration of interest}
The authors declare that they have no known competing financial interests or personal relationships that could have appeared to influence the work reported in this paper.

\bibliographystyle{model5-names}
\bibliography{refs,refs_cited}

\begin{thebibliography}{145}
\expandafter\ifx\csname natexlab\endcsname\relax\def\natexlab#1{#1}\fi
\providecommand{\url}[1]{\texttt{#1}}
\providecommand{\href}[2]{#2}
\providecommand{\path}[1]{#1}
\providecommand{\DOIprefix}{doi:}
\providecommand{\ArXivprefix}{arXiv:}
\providecommand{\URLprefix}{URL: }
\providecommand{\Pubmedprefix}{pmid:}
\providecommand{\doi}[1]{\href{http://dx.doi.org/#1}{\path{#1}}}
\providecommand{\Pubmed}[1]{\href{pmid:#1}{\path{#1}}}
\providecommand{\bibinfo}[2]{#2}
\ifx\xfnm\relax \def\xfnm[#1]{\unskip,\space#1}\fi
\bibitem[{van~der Aalst(2016)}]{van2016data}
\bibinfo{author}{van~der Aalst, W.} (\bibinfo{year}{2016}).
\newblock \bibinfo{title}{Data science in action}.
\newblock In {\it \bibinfo{booktitle}{Process Mining: Data Science in Action}\/} (pp. \bibinfo{pages}{3--23}).
\newblock \bibinfo{address}{Berlin, Heidelberg}: \bibinfo{publisher}{Springer Berlin Heidelberg}.
\newblock \URLprefix \url{https://doi.org/10.1007/978-3-662-49851-4_1}. \DOIprefix\doi{10.1007/978-3-662-49851-4_1}.
\bibitem[{Adderley \& Peterson(2020)}]{adderley2020interactive}
\bibinfo{author}{Adderley, N.}, \& \bibinfo{author}{Peterson, G.} (\bibinfo{year}{2020}).
\newblock \bibinfo{title}{Interactive temporal digital forensic event analysis}.
\newblock In \bibinfo{editor}{G.~Peterson}, \& \bibinfo{editor}{S.~Shenoi} (Eds.), {\it \bibinfo{booktitle}{Advances in {Digital} {Forensics} {XVI}}\/} {IFIP} {Advances} in {Information} and {Communication} {Technology} (pp. \bibinfo{pages}{39--55}).
\newblock \bibinfo{address}{Cham}: \bibinfo{publisher}{Springer International Publishing}.
\newblock \DOIprefix\doi{10.1007/978-3-030-56223-6_3}.
\bibitem[{Adedayo \& Olivier(2015)}]{adedayo2015ideal}
\bibinfo{author}{Adedayo, O.~M.}, \& \bibinfo{author}{Olivier, M.~S.} (\bibinfo{year}{2015}).
\newblock \bibinfo{title}{Ideal log setting for database forensics reconstruction}.
\newblock {\it \bibinfo{journal}{Digital Investigation}\/},  {\it \bibinfo{volume}{12}\/}, \bibinfo{pages}{27--40}.
\bibitem[{Alqahtany et~al.(2016)Alqahtany, Clarke, Furnell \& Reich}]{alqahtany2016forensic}
\bibinfo{author}{Alqahtany, S.}, \bibinfo{author}{Clarke, N.}, \bibinfo{author}{Furnell, S.}, \& \bibinfo{author}{Reich, C.} (\bibinfo{year}{2016}).
\newblock \bibinfo{title}{A forensic acquisition and analysis system for {IaaS}}.
\newblock {\it \bibinfo{journal}{Cluster Computing}\/},  {\it \bibinfo{volume}{19}\/}, \bibinfo{pages}{439--453}. \DOIprefix\doi{10.1007/s10586-015-0509-x}.
\bibitem[{Amato et~al.(2017)Amato, Cozzolino, Mazzeo \& Mazzocca}]{amato2017correlation}
\bibinfo{author}{Amato, F.}, \bibinfo{author}{Cozzolino, G.}, \bibinfo{author}{Mazzeo, A.}, \& \bibinfo{author}{Mazzocca, N.} (\bibinfo{year}{2017}).
\newblock \bibinfo{title}{Correlation of {Digital} {Evidences} in {Forensic} {Investigation} through {Semantic} {Technologies}}.
\newblock In {\it \bibinfo{booktitle}{2017 31st {International} {Conference} on {Advanced} {Information} {Networking} and {Applications} {Workshops} ({WAINA})}\/} (pp. \bibinfo{pages}{668--673}).
\newblock \DOIprefix\doi{10.1109/WAINA.2017.4}.
\bibitem[{Andrade(2020)}]{andrade2020expose}
\bibinfo{author}{Andrade, R.} (\bibinfo{year}{2020}).
\newblock \bibinfo{title}{Expose evidence of timestomping with the ntfs timestamp mismatch artifact}.
\newblock \URLprefix \url{https://www.magnetforensics.com/blog/expose-evidence-of-timestomping-with-the-ntfs-timestamp-mismatch-artifact-in-magnet-axiom-4-4/}.
\bibitem[{Arshad et~al.(2018)Arshad, Jantan \& Abiodun}]{arshad2018digital}
\bibinfo{author}{Arshad, H.}, \bibinfo{author}{Jantan, A.~B.}, \& \bibinfo{author}{Abiodun, O.~I.} (\bibinfo{year}{2018}).
\newblock \bibinfo{title}{Digital forensics: review of issues in scientific validation of digital evidence}.
\newblock {\it \bibinfo{journal}{Journal of Information Processing Systems}\/},  {\it \bibinfo{volume}{14}\/}, \bibinfo{pages}{346--376}.
\bibitem[{Bang et~al.(2009)Bang, Yoo, Kim \& Lee}]{bang2009analysis}
\bibinfo{author}{Bang, J.}, \bibinfo{author}{Yoo, B.}, \bibinfo{author}{Kim, J.}, \& \bibinfo{author}{Lee, S.} (\bibinfo{year}{2009}).
\newblock \bibinfo{title}{Analysis of time information for digital investigation}.
\newblock In {\it \bibinfo{booktitle}{2009 Fifth International Joint Conference on INC, IMS and IDC}\/} (pp. \bibinfo{pages}{1858--1864}).
\newblock \bibinfo{organization}{IEEE}.
\bibitem[{Batten et~al.(2012)Batten, Pan \& Khan}]{batten2012hypothesis}
\bibinfo{author}{Batten, L.}, \bibinfo{author}{Pan, L.}, \& \bibinfo{author}{Khan, N.} (\bibinfo{year}{2012}).
\newblock \bibinfo{title}{Hypothesis generation and testing in event profiling for digital forensic investigations}.
\newblock {\it \bibinfo{journal}{Int. J. Digit. Crime Forensics}\/},  {\it \bibinfo{volume}{4}\/}, \bibinfo{pages}{1--14}. \DOIprefix\doi{10.4018/jdcf.2012100101}.
\bibitem[{Battistoni et~al.(2016)Battistoni, Di~Pietro \& Lombardi}]{battistoni2016cure}
\bibinfo{author}{Battistoni, R.}, \bibinfo{author}{Di~Pietro, R.}, \& \bibinfo{author}{Lombardi, F.} (\bibinfo{year}{2016}).
\newblock \bibinfo{title}{Cure—towards enforcing a reliable timeline for cloud forensics: Model, architecture, and experiments}.
\newblock {\it \bibinfo{journal}{Computer Communications}\/},  {\it \bibinfo{volume}{91}\/}, \bibinfo{pages}{29--43}.
\bibitem[{Becker et~al.(2008)Becker, Rabenseifner \& Wolf}]{becker2008implications}
\bibinfo{author}{Becker, D.}, \bibinfo{author}{Rabenseifner, R.}, \& \bibinfo{author}{Wolf, F.} (\bibinfo{year}{2008}).
\newblock \bibinfo{title}{Implications of non-constant clock drifts for the timestamps of concurrent events}.
\newblock In {\it \bibinfo{booktitle}{2008 IEEE International Conference on Cluster Computing}\/} (pp. \bibinfo{pages}{59--68}).
\bibitem[{Berggren et~al.(2024)Berggren, Gudjonsson, J\"ager et~al.}]{berggren2024timesketch}
\bibinfo{author}{Berggren, J.}, \bibinfo{author}{Gudjonsson, K.}, \bibinfo{author}{J\"ager, A.} et~al. (\bibinfo{year}{2024}).
\newblock \bibinfo{title}{Timesketch: Collaborative forensic timeline analysis}.
\newblock \bibinfo{howpublished}{\url{https://github.com/google/timesketch}}.
\bibitem[{Bhandari \& Jusas(2020)}]{bhandari2020ontology}
\bibinfo{author}{Bhandari, S.}, \& \bibinfo{author}{Jusas, V.} (\bibinfo{year}{2020}).
\newblock \bibinfo{title}{An ontology based on the timeline of {Log2timeline} and {Psort} using abstraction approach in digital forensics}.
\newblock {\it \bibinfo{journal}{Symmetry}\/},  {\it \bibinfo{volume}{12}\/}, \bibinfo{pages}{642}. \URLprefix \url{https://www.mdpi.com/2073-8994/12/4/642}. \DOIprefix\doi{10.3390/sym12040642}.
\newblock \bibinfo{note}{Number: 4 Publisher: Multidisciplinary Digital Publishing Institute}.
\bibitem[{Bhat et~al.(2021)Bhat, AlZahrani \& Wani}]{bhat2021can}
\bibinfo{author}{Bhat, W.~A.}, \bibinfo{author}{AlZahrani, A.}, \& \bibinfo{author}{Wani, M.~A.} (\bibinfo{year}{2021}).
\newblock \bibinfo{title}{Can computer forensic tools be trusted in digital investigations?}
\newblock {\it \bibinfo{journal}{Science \& Justice}\/},  {\it \bibinfo{volume}{61}\/}, \bibinfo{pages}{198--203}.
\bibitem[{Boyd \& Forster(2004)}]{boyd2004time}
\bibinfo{author}{Boyd, C.}, \& \bibinfo{author}{Forster, P.} (\bibinfo{year}{2004}).
\newblock \bibinfo{title}{Time and date issues in forensic computing—a case study}.
\newblock {\it \bibinfo{journal}{Digital Investigation}\/},  {\it \bibinfo{volume}{1}\/}, \bibinfo{pages}{18--23}.
\bibitem[{Breitinger et~al.(2024)Breitinger, Hilgert, Hargreaves, Sheppard, Overdorf \& Scanlon}]{breitinger2024dfrws}
\bibinfo{author}{Breitinger, F.}, \bibinfo{author}{Hilgert, J.-N.}, \bibinfo{author}{Hargreaves, C.}, \bibinfo{author}{Sheppard, J.}, \bibinfo{author}{Overdorf, R.}, \& \bibinfo{author}{Scanlon, M.} (\bibinfo{year}{2024}).
\newblock \bibinfo{title}{Dfrws eu 10-year review and future directions in digital forensic research}.
\newblock {\it \bibinfo{journal}{Forensic Science International: Digital Investigation}\/},  {\it \bibinfo{volume}{48}\/}, \bibinfo{pages}{301685}.
\bibitem[{Buchholz \& Tjaden(2007)}]{buchholz2007brief}
\bibinfo{author}{Buchholz, F.}, \& \bibinfo{author}{Tjaden, B.} (\bibinfo{year}{2007}).
\newblock \bibinfo{title}{A brief study of time}.
\newblock {\it \bibinfo{journal}{Digital Investigation}\/},  {\it \bibinfo{volume}{4}\/}, \bibinfo{pages}{31--42}. \DOIprefix\doi{10.1016/j.diin.2007.06.004}.
\bibitem[{Buchholz \& Falk(2005)}]{buchholz2005design}
\bibinfo{author}{Buchholz, F.~P.}, \& \bibinfo{author}{Falk, C.} (\bibinfo{year}{2005}).
\newblock \bibinfo{title}{Design and implementation of zeitline: a forensic timeline editor.}
\newblock In {\it \bibinfo{booktitle}{DFRWS}\/}.
\bibitem[{Carrier \& Spafford(2004{\natexlab{a}})}]{carrier2004defining}
\bibinfo{author}{Carrier, B.}, \& \bibinfo{author}{Spafford, E.} (\bibinfo{year}{2004}{\natexlab{a}}).
\newblock \bibinfo{title}{Defining event reconstruction of a digital crime scene}.
\newblock {\it \bibinfo{journal}{Journal of Forensic Sciences}\/},  {\it \bibinfo{volume}{49}\/}, \bibinfo{pages}{1291--1298}. \DOIprefix\doi{10.1520/JFS2004127}.
\bibitem[{Carrier \& Spafford(2004{\natexlab{b}})}]{carrier2004event}
\bibinfo{author}{Carrier, B.}, \& \bibinfo{author}{Spafford, E.} (\bibinfo{year}{2004}{\natexlab{b}}).
\newblock \bibinfo{title}{An event-based digital forensic investigation framework}.
\newblock In {\it \bibinfo{booktitle}{Proceedings of the The Digital Forensic Research Conference}\/} (pp. \bibinfo{pages}{1--12}).
\bibitem[{Carrier(2006)}]{carrier2006hypothesis}
\bibinfo{author}{Carrier, B.~D.} (\bibinfo{year}{2006}).
\newblock {\it \bibinfo{title}{A hypothesis-based approach to digital forensic investigations}\/}.
\newblock Ph.D. thesis Purdue University.
\bibitem[{Casey(2011)}]{casey2011digital}
\bibinfo{author}{Casey, E.} (\bibinfo{year}{2011}).
\newblock {\it \bibinfo{title}{Digital evidence and computer crime: forensic science, computers and the {Internet}}\/}.
\newblock (\bibinfo{edition}{3rd} ed.).
\newblock \bibinfo{address}{Waltham, MA}: \bibinfo{publisher}{Academic Press}.
\bibitem[{Casey(2018)}]{casey2018digital}
\bibinfo{author}{Casey, E.} (\bibinfo{year}{2018}).
\newblock \bibinfo{title}{Digital {Stratigraphy}: {Contextual} {Analysis} of {File} {System} {Traces} in {Forensic} {Science}}.
\newblock {\it \bibinfo{journal}{Journal of Forensic Sciences}\/},  {\it \bibinfo{volume}{63}\/}, \bibinfo{pages}{1383--1391}. \DOIprefix\doi{10.1111/1556-4029.13722}.
\newblock \bibinfo{note}{Number: 5}.
\bibitem[{Casey(2020)}]{casey2020standardization}
\bibinfo{author}{Casey, E.} (\bibinfo{year}{2020}).
\newblock \bibinfo{title}{Standardization of forming and expressing preliminary evaluative opinions on digital evidence}.
\newblock {\it \bibinfo{journal}{Forensic Science International: Digital Investigation}\/},  {\it \bibinfo{volume}{32}\/}, \bibinfo{pages}{200888}. \DOIprefix\doi{https://doi.org/10.1016/j.fsidi.2019.200888}.
\bibitem[{Casey et~al.(2022)Casey, Nguyen, Mates \& Lalliss}]{casey2022crowdsourcing}
\bibinfo{author}{Casey, E.}, \bibinfo{author}{Nguyen, L.}, \bibinfo{author}{Mates, J.}, \& \bibinfo{author}{Lalliss, S.} (\bibinfo{year}{2022}).
\newblock \bibinfo{title}{Crowdsourcing forensics: {Creating} a curated catalog of digital forensic artifacts}.
\newblock {\it \bibinfo{journal}{Journal of Forensic Sciences}\/},  {\it \bibinfo{volume}{67}\/}, \bibinfo{pages}{1846--1857}. \DOIprefix\doi{10.1111/1556-4029.15053}.
\newblock \bibinfo{note}{\_eprint: https://onlinelibrary.wiley.com/doi/pdf/10.1111/1556-4029.15053}.
\bibitem[{Chabot et~al.(2014)Chabot, Bertaux, Nicolle \& Kechadi}]{chabot2014complete}
\bibinfo{author}{Chabot, Y.}, \bibinfo{author}{Bertaux, A.}, \bibinfo{author}{Nicolle, C.}, \& \bibinfo{author}{Kechadi, M.-T.} (\bibinfo{year}{2014}).
\newblock \bibinfo{title}{A complete formalized knowledge representation model for advanced digital forensics timeline analysis}.
\newblock {\it \bibinfo{journal}{Digital Investigation}\/},  {\it \bibinfo{volume}{11}\/}, \bibinfo{pages}{S95--S105}. \DOIprefix\doi{10.1016/j.diin.2014.05.009}.
\bibitem[{Chabot et~al.(2015{\natexlab{a}})Chabot, Bertaux, Nicolle \& Kechadi}]{chabot2015event}
\bibinfo{author}{Chabot, Y.}, \bibinfo{author}{Bertaux, A.}, \bibinfo{author}{Nicolle, C.}, \& \bibinfo{author}{Kechadi, M.-T.} (\bibinfo{year}{2015}{\natexlab{a}}).
\newblock \bibinfo{title}{Event {Reconstruction}: {A} {State} of the {Art}}.
\newblock In \bibinfo{editor}{M.~M. Cruz-Cunha}, \bibinfo{editor}{I.~M. Portela}, \& \bibinfo{editor}{A.~Piekarz} (Eds.), {\it \bibinfo{booktitle}{Handbook of {Research} on {Digital} {Crime}, {Cyberspace} {Security}, and {Information} {Assurance}:}\/} Advances in {Digital} {Crime}, {Forensics}, and {Cyber} {Terrorism} (p.~\bibinfo{pages}{15}).
\newblock \bibinfo{publisher}{IGI Global}.
\newblock \DOIprefix\doi{10.4018/978-1-4666-6324-4}.
\bibitem[{Chabot et~al.(2015{\natexlab{b}})Chabot, Bertaux, Nicolle \& Kechadi}]{chabot2015ontology}
\bibinfo{author}{Chabot, Y.}, \bibinfo{author}{Bertaux, A.}, \bibinfo{author}{Nicolle, C.}, \& \bibinfo{author}{Kechadi, T.} (\bibinfo{year}{2015}{\natexlab{b}}).
\newblock \bibinfo{title}{An ontology-based approach for the reconstruction and analysis of digital incidents timelines}.
\newblock {\it \bibinfo{journal}{Digital Investigation}\/},  {\it \bibinfo{volume}{15}\/}, \bibinfo{pages}{83--100}.
\bibitem[{Choi et~al.(2021)Choi, Lee \& Jeong}]{choi2021forensic}
\bibinfo{author}{Choi, H.}, \bibinfo{author}{Lee, S.}, \& \bibinfo{author}{Jeong, D.} (\bibinfo{year}{2021}).
\newblock \bibinfo{title}{Forensic recovery of {SQL} server database: Practical approach}.
\newblock {\it \bibinfo{journal}{IEEE Access}\/},  {\it \bibinfo{volume}{9}\/}, \bibinfo{pages}{14564--14575}.
\bibitem[{Conlan et~al.(2016)Conlan, Baggili \& Breitinger}]{conlan2016anti}
\bibinfo{author}{Conlan, K.}, \bibinfo{author}{Baggili, I.}, \& \bibinfo{author}{Breitinger, F.} (\bibinfo{year}{2016}).
\newblock \bibinfo{title}{Anti-forensics: {Furthering} digital forensic science through a new extended, granular taxonomy}.
\newblock {\it \bibinfo{journal}{Digital Investigation}\/},  {\it \bibinfo{volume}{18}\/}, \bibinfo{pages}{S66--S75}. \DOIprefix\doi{10.1016/j.diin.2016.04.006}.
\bibitem[{Debinski et~al.(2019)Debinski, Breitinger \& Mohan}]{debinski2019timeline2gui}
\bibinfo{author}{Debinski, M.}, \bibinfo{author}{Breitinger, F.}, \& \bibinfo{author}{Mohan, P.} (\bibinfo{year}{2019}).
\newblock \bibinfo{title}{{Timeline2GUI}: {A} {Log2Timeline} {CSV} parser and training scenarios}.
\newblock {\it \bibinfo{journal}{Digital Investigation}\/},  {\it \bibinfo{volume}{28}\/}, \bibinfo{pages}{34--43}. \DOIprefix\doi{10.1016/j.diin.2018.12.004}.
\bibitem[{Dixit et~al.(2018)Dixit, Suriadi, Andrews, Wynn, ter Hofstede, Buijs \& van~der Aalst}]{dixit2018detection}
\bibinfo{author}{Dixit, P.~M.}, \bibinfo{author}{Suriadi, S.}, \bibinfo{author}{Andrews, R.}, \bibinfo{author}{Wynn, M.~T.}, \bibinfo{author}{ter Hofstede, A.~H.}, \bibinfo{author}{Buijs, J.~C.}, \& \bibinfo{author}{van~der Aalst, W.~M.} (\bibinfo{year}{2018}).
\newblock \bibinfo{title}{Detection and interactive repair of event ordering imperfection in process logs}.
\newblock In {\it \bibinfo{booktitle}{Advanced Information Systems Engineering: 30th International Conference, CAiSE 2018, Tallinn, Estonia, June 11-15, 2018, Proceedings 30}\/} (pp. \bibinfo{pages}{274--290}).
\newblock \bibinfo{organization}{Springer}.
\bibitem[{Dreier et~al.(2024)Dreier, Vanini, Hargreaves, Breitinger \& Freiling}]{dreier2024beyond}
\bibinfo{author}{Dreier, L.~M.}, \bibinfo{author}{Vanini, C.}, \bibinfo{author}{Hargreaves, C.~J.}, \bibinfo{author}{Breitinger, F.}, \& \bibinfo{author}{Freiling, F.} (\bibinfo{year}{2024}).
\newblock \bibinfo{title}{Beyond timestamps: Integrating implicit timing information into digital forensic timelines}.
\newblock {\it \bibinfo{journal}{Forensic Science International: Digital Investigation}\/},  {\it \bibinfo{volume}{49}\/}, \bibinfo{pages}{301755}. \DOIprefix\doi{10.1016/j.fsidi.2024.301755}.
\bibitem[{Du et~al.(2020{\natexlab{a}})Du, Hargreaves, Sheppard, Anda, Sayakkara, Le-Khac \& Scanlon}]{du2020sok}
\bibinfo{author}{Du, X.}, \bibinfo{author}{Hargreaves, C.}, \bibinfo{author}{Sheppard, J.}, \bibinfo{author}{Anda, F.}, \bibinfo{author}{Sayakkara, A.}, \bibinfo{author}{Le-Khac, N.-A.}, \& \bibinfo{author}{Scanlon, M.} (\bibinfo{year}{2020}{\natexlab{a}}).
\newblock \bibinfo{title}{{SoK}: Exploring the state of the art and the future potential of artificial intelligence in digital forensic investigation}.
\newblock In {\it \bibinfo{booktitle}{Proceedings of the 15th International Conference on Availability, Reliability and Security}\/} (pp. \bibinfo{pages}{1--10}).
\bibitem[{Du et~al.(2020{\natexlab{b}})Du, Le \& Scanlon}]{du2020automated}
\bibinfo{author}{Du, X.}, \bibinfo{author}{Le, Q.}, \& \bibinfo{author}{Scanlon, M.} (\bibinfo{year}{2020}{\natexlab{b}}).
\newblock \bibinfo{title}{Automated artefact relevancy determination from artefact metadata and associated timeline events}.
\newblock In {\it \bibinfo{booktitle}{2020 International Conference on Cyber Security and Protection of Digital Services (Cyber Security)}\/} (pp. \bibinfo{pages}{1--8}).
\newblock \bibinfo{organization}{IEEE}.
\bibitem[{Fern{\'a}ndez-Fuentes et~al.(2022)Fern{\'a}ndez-Fuentes, Pena \& Cabaleiro}]{fernandez2022digital}
\bibinfo{author}{Fern{\'a}ndez-Fuentes, X.}, \bibinfo{author}{Pena, T.~F.}, \& \bibinfo{author}{Cabaleiro, J.~C.} (\bibinfo{year}{2022}).
\newblock \bibinfo{title}{Digital forensic analysis methodology for private browsing: Firefox and chrome on linux as a case study}.
\newblock {\it \bibinfo{journal}{Computers \& Security}\/},  {\it \bibinfo{volume}{115}\/}, \bibinfo{pages}{102626}.
\bibitem[{Galhuber \& Luh(2021)}]{galhuber2021time}
\bibinfo{author}{Galhuber, M.}, \& \bibinfo{author}{Luh, R.} (\bibinfo{year}{2021}).
\newblock \bibinfo{title}{Time for {Truth}: {Forensic} {Analysis} of {NTFS} {Timestamps}}.
\newblock In {\it \bibinfo{booktitle}{Proceedings of the 16th {International} {Conference} on {Availability}, {Reliability} and {Security}}\/} {ARES} 21 (pp. \bibinfo{pages}{1--10}).
\newblock \bibinfo{address}{New York, NY, USA}: \bibinfo{publisher}{Association for Computing Machinery}.
\newblock \DOIprefix\doi{10.1145/3465481.3470016}.
\bibitem[{Gladyshev \& Patel(2004)}]{gladyshev2004finite}
\bibinfo{author}{Gladyshev, P.}, \& \bibinfo{author}{Patel, A.} (\bibinfo{year}{2004}).
\newblock \bibinfo{title}{Finite state machine approach to digital event reconstruction}.
\newblock {\it \bibinfo{journal}{Digital Investigation}\/},  {\it \bibinfo{volume}{1}\/}, \bibinfo{pages}{130--149}. \DOIprefix\doi{10.1016/j.diin.2004.03.001}.
\bibitem[{G{\'o}mez et~al.(2005)G{\'o}mez, Herrerias \& Mata}]{gomez2005using}
\bibinfo{author}{G{\'o}mez, R.}, \bibinfo{author}{Herrerias, J.}, \& \bibinfo{author}{Mata, E.} (\bibinfo{year}{2005}).
\newblock \bibinfo{title}{Using lamport’s logical clocks to consolidate log files from different sources}.
\newblock In {\it \bibinfo{booktitle}{International Workshop on Innovative Internet Community Systems}\/} (pp. \bibinfo{pages}{126--133}).
\newblock \bibinfo{organization}{Springer}.
\bibitem[{Grajeda et~al.(2017)Grajeda, Breitinger \& Baggili}]{grajeda2017availability}
\bibinfo{author}{Grajeda, C.}, \bibinfo{author}{Breitinger, F.}, \& \bibinfo{author}{Baggili, I.} (\bibinfo{year}{2017}).
\newblock \bibinfo{title}{Availability of datasets for digital forensics--and what is missing}.
\newblock {\it \bibinfo{journal}{Digital Investigation}\/},  {\it \bibinfo{volume}{22}\/}, \bibinfo{pages}{S94--S105}.
\bibitem[{Grajeda et~al.(2018)Grajeda, Sanchez, Baggili, Clark \& Breitinger}]{grajeda2018experience}
\bibinfo{author}{Grajeda, C.}, \bibinfo{author}{Sanchez, L.}, \bibinfo{author}{Baggili, I.}, \bibinfo{author}{Clark, D.}, \& \bibinfo{author}{Breitinger, F.} (\bibinfo{year}{2018}).
\newblock \bibinfo{title}{Experience constructing the artifact genome project (agp): managing the domain's knowledge one artifact at a time}.
\newblock {\it \bibinfo{journal}{Digital Investigation}\/},  {\it \bibinfo{volume}{26}\/}, \bibinfo{pages}{S47--S58}.
\bibitem[{Group et~al.(2014)}]{nist2014nist}
\bibinfo{author}{Group, N. C. C. F. S.~W.} et~al. (\bibinfo{year}{2014}).
\newblock {\it \bibinfo{title}{{NIST} cloud computing forensic science challenges}\/}.
\newblock \bibinfo{type}{Technical Report} National Institute of Standards and Technology.
\bibitem[{Gruber et~al.(2023)Gruber, Hargreaves \& Freiling}]{gruber2023contamination}
\bibinfo{author}{Gruber, J.}, \bibinfo{author}{Hargreaves, C.~J.}, \& \bibinfo{author}{Freiling, F.~C.} (\bibinfo{year}{2023}).
\newblock \bibinfo{title}{Contamination of digital evidence: {Understanding} an underexposed risk}.
\newblock {\it \bibinfo{journal}{Forensic Science International: Digital Investigation}\/},  {\it \bibinfo{volume}{44}\/}, \bibinfo{pages}{301501}. \DOIprefix\doi{10.1016/j.fsidi.2023.301501}.
\bibitem[{Gu{\dh}j{\'o}nsson(2010)}]{gudhjonsson2010mastering}
\bibinfo{author}{Gu{\dh}j{\'o}nsson, K.} (\bibinfo{year}{2010}).
\newblock \bibinfo{title}{Mastering the super timeline with log2timeline}.
\newblock {\it \bibinfo{journal}{SANS Institute}\/}, .
\bibitem[{Hampton \& Baig(2016)}]{hampton2016timestamp}
\bibinfo{author}{Hampton, N.}, \& \bibinfo{author}{Baig, Z.~A.} (\bibinfo{year}{2016}).
\newblock \bibinfo{title}{Timestamp analysis for quality validation of network forensic data}.
\newblock In {\it \bibinfo{booktitle}{Network and System Security: 10th International Conference, NSS 2016, Taipei, Taiwan, September 28-30, 2016, Proceedings 10}\/} (pp. \bibinfo{pages}{235--248}).
\newblock \bibinfo{organization}{Springer}.
\bibitem[{Han et~al.(2020)Han, Kim \& Lee}]{han20205w1h}
\bibinfo{author}{Han, J.}, \bibinfo{author}{Kim, J.}, \& \bibinfo{author}{Lee, S.} (\bibinfo{year}{2020}).
\newblock \bibinfo{title}{5w1h-based expression for the effective sharing of information in digital forensic investigations}.
\newblock {\it \bibinfo{journal}{arXiv preprint arXiv:2010.15711}\/}, .
\bibitem[{Hargreaves et~al.(2025)Hargreaves, van Beek \& Casey}]{hargreaves2025solve}
\bibinfo{author}{Hargreaves, C.}, \bibinfo{author}{van Beek, H.}, \& \bibinfo{author}{Casey, E.} (\bibinfo{year}{2025}).
\newblock \bibinfo{title}{Solve-it: A proposed digital forensic knowledge base inspired by mitre att\&ck}.
\newblock {\it \bibinfo{journal}{Forensic Science International: Digital Investigation}\/},  {\it \bibinfo{volume}{52}\/}, \bibinfo{pages}{301864}.
\bibitem[{Hargreaves et~al.(2024{\natexlab{a}})Hargreaves, Breitinger, Dowthwaite, Webb \& Scanlon}]{hargreaves2024dfpulse}
\bibinfo{author}{Hargreaves, C.}, \bibinfo{author}{Breitinger, F.}, \bibinfo{author}{Dowthwaite, L.}, \bibinfo{author}{Webb, H.}, \& \bibinfo{author}{Scanlon, M.} (\bibinfo{year}{2024}{\natexlab{a}}).
\newblock \bibinfo{title}{Dfpulse: The 2024 digital forensic practitioner survey}.
\newblock {\it \bibinfo{journal}{Forensic Science International: Digital Investigation}\/},  {\it \bibinfo{volume}{51}\/}, \bibinfo{pages}{301844}.
\bibitem[{Hargreaves et~al.(2024{\natexlab{b}})Hargreaves, Nelson \& Casey}]{hargreaves2024abstract}
\bibinfo{author}{Hargreaves, C.}, \bibinfo{author}{Nelson, A.}, \& \bibinfo{author}{Casey, E.} (\bibinfo{year}{2024}{\natexlab{b}}).
\newblock \bibinfo{title}{An abstract model for digital forensic analysis tools-a foundation for systematic error mitigation analysis}.
\newblock {\it \bibinfo{journal}{Forensic Science International: Digital Investigation}\/},  {\it \bibinfo{volume}{48}\/}, \bibinfo{pages}{301679}. \DOIprefix\doi{10.1016/j.fsidi.2023.301679}.
\bibitem[{Hargreaves \& Patterson(2012)}]{hargreaves2012automated}
\bibinfo{author}{Hargreaves, C.}, \& \bibinfo{author}{Patterson, J.} (\bibinfo{year}{2012}).
\newblock \bibinfo{title}{An automated timeline reconstruction approach for digital forensic investigations}.
\newblock {\it \bibinfo{journal}{Digital Investigation}\/},  {\it \bibinfo{volume}{9}\/}, \bibinfo{pages}{S69--S79}. \DOIprefix\doi{10.1016/j.diin.2012.05.006}.
\bibitem[{Hargreaves(2009)}]{hargreaves2009assessing}
\bibinfo{author}{Hargreaves, C.~J.} (\bibinfo{year}{2009}).
\newblock {\it \bibinfo{title}{Assessing the reliability of digital evidence from live investigations involving encryption.}\/}.
\newblock Ph.D. thesis Cranfield University, UK.
\bibitem[{Harichandran et~al.(2016)Harichandran, Walnycky, Baggili \& Breitinger}]{harichandran2016cufa}
\bibinfo{author}{Harichandran, V.~S.}, \bibinfo{author}{Walnycky, D.}, \bibinfo{author}{Baggili, I.}, \& \bibinfo{author}{Breitinger, F.} (\bibinfo{year}{2016}).
\newblock \bibinfo{title}{Cufa: A more formal definition for digital forensic artifacts}.
\newblock {\it \bibinfo{journal}{Digital Investigation}\/},  {\it \bibinfo{volume}{18}\/}, \bibinfo{pages}{S125--S137}.
\bibitem[{Henderson(2009)}]{henderson2009categorization}
\bibinfo{author}{Henderson, G.} (\bibinfo{year}{2009}).
\newblock {\it \bibinfo{title}{A Categorization of Computer Clocks}\/}.
\newblock \bibinfo{type}{Technical Report} Department of Computer Science, James Madison University.
\bibitem[{Henseler \& van Beek(2023)}]{henseler2023chatgpt}
\bibinfo{author}{Henseler, H.}, \& \bibinfo{author}{van Beek, H.} (\bibinfo{year}{2023}).
\newblock \bibinfo{title}{Chatgpt as a copilot for investigating digital evidence.}
\newblock In {\it \bibinfo{booktitle}{Proceedings of the Third International Workshop on Artificial Intelligence and Intelligent Assistance for Legal Professionals in the Digital Workplace}\/} (pp. \bibinfo{pages}{58--69}).
\bibitem[{Horsman(2018)}]{horsman2018couldn}
\bibinfo{author}{Horsman, G.} (\bibinfo{year}{2018}).
\newblock \bibinfo{title}{“i couldn't find it your honour, it mustn't be there!”--tool errors, tool limitations and user error in digital forensics}.
\newblock {\it \bibinfo{journal}{Science \& Justice}\/},  {\it \bibinfo{volume}{58}\/}, \bibinfo{pages}{433--440}.
\bibitem[{Horsman(2019)}]{horsman2019raiders}
\bibinfo{author}{Horsman, G.} (\bibinfo{year}{2019}).
\newblock \bibinfo{title}{Raiders of the lost artefacts: Championing the need for digital forensics research}.
\newblock {\it \bibinfo{journal}{Forensic Science International: Reports}\/},  {\it \bibinfo{volume}{1}\/}, \bibinfo{pages}{100003}.
\bibitem[{Inglot \& Liu(2014)}]{inglot2014enhanced}
\bibinfo{author}{Inglot, B.}, \& \bibinfo{author}{Liu, L.} (\bibinfo{year}{2014}).
\newblock \bibinfo{title}{Enhanced timeline analysis for digital forensic investigations}.
\newblock {\it \bibinfo{journal}{Information Security Journal: A Global Perspective}\/},  {\it \bibinfo{volume}{23}\/}, \bibinfo{pages}{32--44}.
\bibitem[{Jahankhani \& Hosseinian-far(2014)}]{jahankhani2014digital}
\bibinfo{author}{Jahankhani, H.}, \& \bibinfo{author}{Hosseinian-far, A.} (\bibinfo{year}{2014}).
\newblock \bibinfo{title}{Digital forensics education, training and awareness}.
\newblock In {\it \bibinfo{booktitle}{Cyber Crime and Cyber Terrorism Investigator's Handbook}\/} (pp. \bibinfo{pages}{91--100}).
\newblock \bibinfo{publisher}{Elsevier}.
\newblock \DOIprefix\doi{10.1016/B978-0-12-800743-3.00008-6}.
\bibitem[{Jang et~al.(2016)Jang, Ahn, Hwang \& Kim}]{jang2016understanding}
\bibinfo{author}{Jang, D.-i.}, \bibinfo{author}{Ahn, G.-J.}, \bibinfo{author}{Hwang, H.}, \& \bibinfo{author}{Kim, K.} (\bibinfo{year}{2016}).
\newblock \bibinfo{title}{Understanding anti-forensic techniques with timestamp manipulation}.
\newblock In {\it \bibinfo{booktitle}{2016 IEEE 17th International Conference on Information Reuse and Integration (IRI)}\/} (pp. \bibinfo{pages}{609--614}).
\newblock \bibinfo{organization}{IEEE}.
\bibitem[{Jaquet-Chiffelle \& Casey(2021)}]{jaquet2021formalized}
\bibinfo{author}{Jaquet-Chiffelle, D.-O.}, \& \bibinfo{author}{Casey, E.} (\bibinfo{year}{2021}).
\newblock \bibinfo{title}{A formalized model of the {Trace}}.
\newblock {\it \bibinfo{journal}{Forensic Science International}\/},  {\it \bibinfo{volume}{327}\/}, \bibinfo{pages}{110941}. \DOIprefix\doi{10.1016/j.forsciint.2021.110941}.
\bibitem[{Jarrett \& Choo(2021)}]{jarrett2021impact}
\bibinfo{author}{Jarrett, A.}, \& \bibinfo{author}{Choo, K.-K.~R.} (\bibinfo{year}{2021}).
\newblock \bibinfo{title}{The impact of automation and artificial intelligence on digital forensics}.
\newblock {\it \bibinfo{journal}{Wiley Interdisciplinary Reviews: Forensic Science}\/},  {\it \bibinfo{volume}{3}\/}, \bibinfo{pages}{e1418}.
\bibitem[{Jinad et~al.(2024)Jinad, Gupta, Simsek \& Zhou}]{jinad2024bias}
\bibinfo{author}{Jinad, R.}, \bibinfo{author}{Gupta, K.}, \bibinfo{author}{Simsek, E.}, \& \bibinfo{author}{Zhou, B.} (\bibinfo{year}{2024}).
\newblock \bibinfo{title}{Bias and fairness in software and automation tools in digital forensics}.
\newblock {\it \bibinfo{journal}{J. Surveill. Secur. Saf}\/},  {\it \bibinfo{volume}{5}\/}, \bibinfo{pages}{19--35}.
\bibitem[{Joseph \& Singh(2019)}]{joseph2019digital}
\bibinfo{author}{Joseph, A.}, \& \bibinfo{author}{Singh, K.~J.} (\bibinfo{year}{2019}).
\newblock \bibinfo{title}{Digital forensics in distributed environment}.
\newblock In {\it \bibinfo{booktitle}{Cloud Security: Concepts, Methodologies, Tools, and Applications}\/} (pp. \bibinfo{pages}{1157--1177}).
\newblock \bibinfo{publisher}{IGI Global}.
\bibitem[{J{\"u}rgensen(2021)}]{jurgensen2021trace}
\bibinfo{author}{J{\"u}rgensen, J.~P.} (\bibinfo{year}{2021}).
\newblock \bibinfo{title}{Trace reconstruction in system logs for processing with process mining}.
\newblock {\it \bibinfo{journal}{Procedia Computer Science}\/},  {\it \bibinfo{volume}{180}\/}, \bibinfo{pages}{352--357}.
\bibitem[{Kaart \& Laraghy(2014)}]{kaart2014android}
\bibinfo{author}{Kaart, M.}, \& \bibinfo{author}{Laraghy, S.} (\bibinfo{year}{2014}).
\newblock \bibinfo{title}{Android forensics: Interpretation of timestamps}.
\newblock {\it \bibinfo{journal}{Digital Investigation}\/},  {\it \bibinfo{volume}{11}\/}, \bibinfo{pages}{234--248}. \DOIprefix\doi{10.1016/j.diin.2014.05.001}.
\bibitem[{Kang et~al.(2013)Kang, Lee \& Lee}]{kang2013digital}
\bibinfo{author}{Kang, J.}, \bibinfo{author}{Lee, S.}, \& \bibinfo{author}{Lee, H.} (\bibinfo{year}{2013}).
\newblock \bibinfo{title}{A digital forensic framework for automated user activity reconstruction}.
\newblock In {\it \bibinfo{booktitle}{Information Security Practice and Experience: 9th International Conference, ISPEC 2013, Lanzhou, China, May 12-14, 2013. Proceedings 9}\/} (pp. \bibinfo{pages}{263--277}).
\newblock \bibinfo{organization}{Springer}.
\bibitem[{Karie \& Venter(2015)}]{karie2015taxonomy}
\bibinfo{author}{Karie, N.~M.}, \& \bibinfo{author}{Venter, H.~S.} (\bibinfo{year}{2015}).
\newblock \bibinfo{title}{Taxonomy of challenges for digital forensics}.
\newblock {\it \bibinfo{journal}{Journal of Forensic Sciences}\/},  {\it \bibinfo{volume}{60}\/}, \bibinfo{pages}{885--893}.
\bibitem[{Kassin et~al.(2013)Kassin, Dror \& Kukucka}]{kassin2013forensic}
\bibinfo{author}{Kassin, S.~M.}, \bibinfo{author}{Dror, I.~E.}, \& \bibinfo{author}{Kukucka, J.} (\bibinfo{year}{2013}).
\newblock \bibinfo{title}{The forensic confirmation bias: Problems, perspectives, and proposed solutions}.
\newblock {\it \bibinfo{journal}{Journal of applied research in memory and cognition}\/},  {\it \bibinfo{volume}{2}\/}, \bibinfo{pages}{42--52}.
\bibitem[{Kebande \& Venter(2018)}]{kebande2018novel}
\bibinfo{author}{Kebande, V.~R.}, \& \bibinfo{author}{Venter, H.~S.} (\bibinfo{year}{2018}).
\newblock \bibinfo{title}{Novel digital forensic readiness technique in the cloud environment}.
\newblock {\it \bibinfo{journal}{Australian Journal of Forensic Sciences}\/},  {\it \bibinfo{volume}{50}\/}, \bibinfo{pages}{552--591}. \DOIprefix\doi{10.1080/00450618.2016.1267797}.
\bibitem[{Khan et~al.(2007)Khan, Chatwin \& Young}]{khan2007framework}
\bibinfo{author}{Khan, M.}, \bibinfo{author}{Chatwin, C.~R.}, \& \bibinfo{author}{Young, R.~C.} (\bibinfo{year}{2007}).
\newblock \bibinfo{title}{A framework for post-event timeline reconstruction using neural networks}.
\newblock {\it \bibinfo{journal}{Digital Investigation}\/},  {\it \bibinfo{volume}{4}\/}, \bibinfo{pages}{146--157}.
\bibitem[{Khan \& Wakeman(2006)}]{khan2006machine}
\bibinfo{author}{Khan, M.}, \& \bibinfo{author}{Wakeman, I.} (\bibinfo{year}{2006}).
\newblock \bibinfo{title}{Machine learning for post-event timeline reconstruction}.
\newblock In {\it \bibinfo{booktitle}{First Conference on Advances in Computer Security and Forensics, Liverpool, UK}\/}.
\newblock \bibinfo{organization}{Citeseer}.
\bibitem[{Khan et~al.(2016)Khan, Gani, Wahab, Bagiwa, Shiraz, Khan, Buyya \& Zomaya}]{khan2016cloud}
\bibinfo{author}{Khan, S.}, \bibinfo{author}{Gani, A.}, \bibinfo{author}{Wahab, A. W.~A.}, \bibinfo{author}{Bagiwa, M.~A.}, \bibinfo{author}{Shiraz, M.}, \bibinfo{author}{Khan, S.~U.}, \bibinfo{author}{Buyya, R.}, \& \bibinfo{author}{Zomaya, A.~Y.} (\bibinfo{year}{2016}).
\newblock \bibinfo{title}{Cloud log forensics: Foundations, state of the art, and future directions}.
\newblock {\it \bibinfo{journal}{ACM Computing Surveys (CSUR)}\/},  {\it \bibinfo{volume}{49}\/}, \bibinfo{pages}{1--42}. \DOIprefix\doi{10.1145/2906149}.
\bibitem[{Kiernan \& Terzi(2009)}]{kiernan2009eventsummarizer}
\bibinfo{author}{Kiernan, J.}, \& \bibinfo{author}{Terzi, E.} (\bibinfo{year}{2009}).
\newblock \bibinfo{title}{Eventsummarizer: a tool for summarizing large event sequences}.
\newblock In {\it \bibinfo{booktitle}{Proceedings of the 12th International Conference on Extending Database Technology: Advances in Database Technology}\/} (pp. \bibinfo{pages}{1136--1139}).
\bibitem[{K{\l}os \& El~Fray(2020)}]{klos2020securing}
\bibinfo{author}{K{\l}os, M.}, \& \bibinfo{author}{El~Fray, I.} (\bibinfo{year}{2020}).
\newblock \bibinfo{title}{Securing event logs with blockchain for iot}.
\newblock In {\it \bibinfo{booktitle}{International Conference on Computer Information Systems and Industrial Management}\/} (pp. \bibinfo{pages}{77--87}).
\newblock \bibinfo{organization}{Springer}.
\newblock \DOIprefix\doi{10.1007/978-3-030-47679-3_7}.
\bibitem[{Kwan et~al.(2008)Kwan, Chow, Law \& Lai}]{kwan2008reasoning}
\bibinfo{author}{Kwan, M.}, \bibinfo{author}{Chow, K.-P.}, \bibinfo{author}{Law, F.}, \& \bibinfo{author}{Lai, P.} (\bibinfo{year}{2008}).
\newblock \bibinfo{title}{Reasoning about evidence using bayesian networks}.
\newblock In {\it \bibinfo{booktitle}{Advances in Digital Forensics IV 4}\/} (pp. \bibinfo{pages}{275--289}).
\newblock \bibinfo{organization}{Springer}.
\bibitem[{Kälber et~al.(2013)Kälber, Dewald \& Freiling}]{kalber2013forensic}
\bibinfo{author}{Kälber, S.}, \bibinfo{author}{Dewald, A.}, \& \bibinfo{author}{Freiling, F.~C.} (\bibinfo{year}{2013}).
\newblock \bibinfo{title}{Forensic {Application}-{Fingerprinting} {Based} on {File} {System} {Metadata}}.
\newblock In {\it \bibinfo{booktitle}{2013 {Seventh} {International} {Conference} on {IT} {Security} {Incident} {Management} and {IT} {Forensics}}\/} (pp. \bibinfo{pages}{98--112}).
\newblock \DOIprefix\doi{10.1109/IMF.2013.20}.
\bibitem[{Latzo \& Freiling(2019)}]{latzo2019characterizing}
\bibinfo{author}{Latzo, T.}, \& \bibinfo{author}{Freiling, F.} (\bibinfo{year}{2019}).
\newblock \bibinfo{title}{Characterizing the {Limitations} of {Forensic} {Event} {Reconstruction} {Based} on {Log} {Files}}.
\newblock In {\it \bibinfo{booktitle}{2019 18th {IEEE} {International} {Conference} {On} {Trust}, {Security} {And} {Privacy} {In} {Computing} {And} {Communications}/13th {IEEE} {International} {Conference} {On} {Big} {Data} {Science} {And} {Engineering} ({TrustCom}/{BigDataSE})}\/} (pp. \bibinfo{pages}{466--475}).
\newblock \DOIprefix\doi{10.1109/TrustCom/BigDataSE.2019.00069} \bibinfo{note}{iSSN: 2324-9013}.
\bibitem[{Lee et~al.(2001)Lee, Palmbach \& Miller}]{lee2001henry}
\bibinfo{author}{Lee, H.~C.}, \bibinfo{author}{Palmbach, T.}, \& \bibinfo{author}{Miller, M.~T.} (\bibinfo{year}{2001}).
\newblock {\it \bibinfo{title}{Henry Lee's crime scene handbook}\/}.
\newblock \bibinfo{publisher}{Academic Press}.
\bibitem[{Levett et~al.(2010)Levett, Jhumka \& Anand}]{levett2010towards}
\bibinfo{author}{Levett, C.~P.}, \bibinfo{author}{Jhumka, A.}, \& \bibinfo{author}{Anand, S.~S.} (\bibinfo{year}{2010}).
\newblock \bibinfo{title}{Towards event ordering in digital forensics}.
\newblock In {\it \bibinfo{booktitle}{Proceedings of the 12th ACM workshop on Multimedia and security}\/} (pp. \bibinfo{pages}{35--42}).
\bibitem[{Losavio et~al.(2015)Losavio, Pastukov \& Polyakova}]{losavio2015cyber}
\bibinfo{author}{Losavio, M.}, \bibinfo{author}{Pastukov, P.}, \& \bibinfo{author}{Polyakova, S.} (\bibinfo{year}{2015}).
\newblock \bibinfo{title}{Cyber black box/event data recorder: legal and ethical perspectives and challenges with digital forensics}.
\newblock {\it \bibinfo{journal}{Journal of Digital Forensics, Security and Law}\/},  {\it \bibinfo{volume}{10}\/}, \bibinfo{pages}{4}.
\bibitem[{Lyle et~al.(2022)Lyle, Guttman, Butler, Sauerwein, Reed \& Lloyd}]{lyle2022digital}
\bibinfo{author}{Lyle, J.~R.}, \bibinfo{author}{Guttman, B.}, \bibinfo{author}{Butler, J.~M.}, \bibinfo{author}{Sauerwein, K.}, \bibinfo{author}{Reed, C.}, \& \bibinfo{author}{Lloyd, C.~E.} (\bibinfo{year}{2022}).
\newblock {\it \bibinfo{title}{Digital {Investigation} {Techniques}: {A} {NIST} {Scientific} {Foundation} {Review}}\/}.
\newblock \bibinfo{type}{Technical Report} National Institute of Standards and Technology.
\newblock \DOIprefix\doi{10.6028/NIST.IR.8354-draft}.
\bibitem[{Malhotra et~al.(2015)Malhotra, Cohen, Brakke \& Goldberg}]{malhotra2015attacking}
\bibinfo{author}{Malhotra, A.}, \bibinfo{author}{Cohen, I.~E.}, \bibinfo{author}{Brakke, E.}, \& \bibinfo{author}{Goldberg, S.} (\bibinfo{year}{2015}).
\newblock \bibinfo{title}{Attacking the network time protocol}.
\newblock {\it \bibinfo{journal}{Cryptology ePrint Archive}\/}, .
\bibitem[{Manral et~al.(2019)Manral, Somani, Choo, Conti \& Gaur}]{manral2019systematic}
\bibinfo{author}{Manral, B.}, \bibinfo{author}{Somani, G.}, \bibinfo{author}{Choo, K.-K.~R.}, \bibinfo{author}{Conti, M.}, \& \bibinfo{author}{Gaur, M.~S.} (\bibinfo{year}{2019}).
\newblock \bibinfo{title}{A systematic survey on cloud forensics challenges, solutions, and future directions}.
\newblock {\it \bibinfo{journal}{ACM Computing Surveys (CSUR)}\/},  {\it \bibinfo{volume}{52}\/}, \bibinfo{pages}{1--38}.
\bibitem[{Marangos et~al.(2016)Marangos, Rizomiliotis \& Mitrou}]{marangos2016time}
\bibinfo{author}{Marangos, N.}, \bibinfo{author}{Rizomiliotis, P.}, \& \bibinfo{author}{Mitrou, L.} (\bibinfo{year}{2016}).
\newblock \bibinfo{title}{Time synchronization: pivotal element in cloud forensics}.
\newblock {\it \bibinfo{journal}{Security and Communication Networks}\/},  {\it \bibinfo{volume}{9}\/}, \bibinfo{pages}{571--582}.
\bibitem[{Markov{\'a} et~al.(2022)Markov{\'a}, Sokol \& Kov{\'a}{\'c}ov{\'a}}]{markova2022detection}
\bibinfo{author}{Markov{\'a}, E.}, \bibinfo{author}{Sokol, P.}, \& \bibinfo{author}{Kov{\'a}{\'c}ov{\'a}, K.} (\bibinfo{year}{2022}).
\newblock \bibinfo{title}{Detection of relevant digital evidence in the forensic timelines}.
\newblock In {\it \bibinfo{booktitle}{2022 14th International Conference on Electronics, Computers and Artificial Intelligence (ECAI)}\/} (pp. \bibinfo{pages}{1--7}).
\newblock \bibinfo{organization}{IEEE}.
\bibitem[{Marrington et~al.(2011)Marrington, Baggili, Mohay \& Clark}]{marrington2011cat}
\bibinfo{author}{Marrington, A.}, \bibinfo{author}{Baggili, I.}, \bibinfo{author}{Mohay, G.}, \& \bibinfo{author}{Clark, A.} (\bibinfo{year}{2011}).
\newblock \bibinfo{title}{{CAT Detect (Computer Activity Timeline Detection)}: A tool for detecting inconsistency in computer activity timelines}.
\newblock {\it \bibinfo{journal}{Digital Investigation}\/},  {\it \bibinfo{volume}{8}\/}, \bibinfo{pages}{S52--S61}.
\bibitem[{Marrington et~al.(2007)Marrington, Mohay, Clark \& Morarji}]{marrington2007event}
\bibinfo{author}{Marrington, A.}, \bibinfo{author}{Mohay, G.}, \bibinfo{author}{Clark, A.}, \& \bibinfo{author}{Morarji, H.} (\bibinfo{year}{2007}).
\newblock \bibinfo{title}{Event-based computer profiling for the forensic reconstruction of computer activity}.
\newblock {\it \bibinfo{journal}{AusCERT 2007, IT-Security: Finding the Balance}\/},  (pp. \bibinfo{pages}{71--87}).
\bibitem[{Metz et~al.(2024)Metz, Gudjonsson, White et~al.}]{metz2024log2timeline}
\bibinfo{author}{Metz, J.}, \bibinfo{author}{Gudjonsson, K.}, \bibinfo{author}{White, D.} et~al. (\bibinfo{year}{2024}).
\newblock \bibinfo{title}{{log2timeline Plaso: Super timeline all the things}}.
\newblock \bibinfo{howpublished}{\url{https://github.com/log2timeline/plaso}}.
\bibitem[{Michelet \& Breitinger(2024)}]{michelet2024chatgpt}
\bibinfo{author}{Michelet, G.}, \& \bibinfo{author}{Breitinger, F.} (\bibinfo{year}{2024}).
\newblock \bibinfo{title}{Chatgpt, llama, can you write my report? an experiment on assisted digital forensics reports written using (local) large language models}.
\newblock {\it \bibinfo{journal}{Forensic Science International: Digital Investigation}\/},  {\it \bibinfo{volume}{48}\/}, \bibinfo{pages}{301683}.
\bibitem[{{MITRE}(2023)}]{mitre2023impair}
\bibinfo{author}{{MITRE}} (\bibinfo{year}{2023}).
\newblock \bibinfo{title}{Impair defenses}.
\newblock \bibinfo{howpublished}{\url{https://attack.mitre.org/techniques/T1562/}}.
\bibitem[{Mohammed et~al.(2016)Mohammed, Clarke \& Li}]{mohammed2016automated}
\bibinfo{author}{Mohammed, H.}, \bibinfo{author}{Clarke, N.}, \& \bibinfo{author}{Li, F.} (\bibinfo{year}{2016}).
\newblock \bibinfo{title}{An automated approach for digital forensic analysis of heterogeneous big data}.
\newblock {\it \bibinfo{journal}{Journal of Digital Forensics, Security and Law}\/},  {\it \bibinfo{volume}{11}\/}, \bibinfo{pages}{9}.
\bibitem[{Neale(2023)}]{neale2023fool}
\bibinfo{author}{Neale, C.} (\bibinfo{year}{2023}).
\newblock \bibinfo{title}{Fool me once: {A} systematic review of techniques to authenticate digital artefacts}.
\newblock {\it \bibinfo{journal}{Forensic Science International: Digital Investigation}\/},  {\it \bibinfo{volume}{45}\/}, \bibinfo{pages}{301516}. \DOIprefix\doi{10.1016/j.fsidi.2023.301516}.
\bibitem[{Neale et~al.(2022)Neale, Kennedy, Price, Yu \& Nuseibeh}]{neale2022case}
\bibinfo{author}{Neale, C.}, \bibinfo{author}{Kennedy, I.}, \bibinfo{author}{Price, B.}, \bibinfo{author}{Yu, Y.}, \& \bibinfo{author}{Nuseibeh, B.} (\bibinfo{year}{2022}).
\newblock \bibinfo{title}{The case for zero trust digital forensics}.
\newblock {\it \bibinfo{journal}{Forensic Science International: Digital Investigation}\/},  {\it \bibinfo{volume}{40}\/}, \bibinfo{pages}{301352}.
\bibitem[{Neasbitt et~al.(2014)Neasbitt, Perdisci, Li \& Nelms}]{neasbitt2014clickminer}
\bibinfo{author}{Neasbitt, C.}, \bibinfo{author}{Perdisci, R.}, \bibinfo{author}{Li, K.}, \& \bibinfo{author}{Nelms, T.} (\bibinfo{year}{2014}).
\newblock \bibinfo{title}{Clickminer: Towards forensic reconstruction of user-browser interactions from network traces}.
\newblock In {\it \bibinfo{booktitle}{Proceedings of the 2014 ACM SIGSAC Conference on Computer and Communications Security}\/} (pp. \bibinfo{pages}{1244--1255}).
\bibitem[{Nguyen \& Comuzzi(2019)}]{nguyen2019event}
\bibinfo{author}{Nguyen, H. T.~C.}, \& \bibinfo{author}{Comuzzi, M.} (\bibinfo{year}{2019}).
\newblock \bibinfo{title}{Event log reconstruction using autoencoders}.
\newblock In {\it \bibinfo{booktitle}{Service-Oriented Computing--ICSOC 2018 Workshops}\/} (pp. \bibinfo{pages}{335--350}).
\newblock \bibinfo{organization}{Springer}.
\bibitem[{Nordvik \& Axelsson(2022)}]{nordvik2022time}
\bibinfo{author}{Nordvik, R.}, \& \bibinfo{author}{Axelsson, S.} (\bibinfo{year}{2022}).
\newblock \bibinfo{title}{It is about time--do exfat implementations handle timestamps correctly?}
\newblock {\it \bibinfo{journal}{Forensic Science International: Digital Investigation}\/},  {\it \bibinfo{volume}{42}\/}, \bibinfo{pages}{301476}.
\bibitem[{Oh et~al.(2022)Oh, Lee \& Hwang}]{oh2022forensic}
\bibinfo{author}{Oh, J.}, \bibinfo{author}{Lee, S.}, \& \bibinfo{author}{Hwang, H.} (\bibinfo{year}{2022}).
\newblock \bibinfo{title}{Forensic recovery of file system metadata for digital forensic investigation}.
\newblock {\it \bibinfo{journal}{IEEE Access}\/},  {\it \bibinfo{volume}{10}\/}, \bibinfo{pages}{111591--111606}.
\bibitem[{Osborne \& Turnbull(2009)}]{osborne2009enhancing}
\bibinfo{author}{Osborne, G.}, \& \bibinfo{author}{Turnbull, B.} (\bibinfo{year}{2009}).
\newblock \bibinfo{title}{Enhancing computer forensics investigation through visualisation and data exploitation}.
\newblock In {\it \bibinfo{booktitle}{2009 International Conference on Availability, Reliability and Security}\/} (pp. \bibinfo{pages}{1012--1017}).
\newblock \bibinfo{organization}{IEEE}.
\bibitem[{Palmbach \& Breitinger(2020)}]{palmbach2020artifacts}
\bibinfo{author}{Palmbach, D.}, \& \bibinfo{author}{Breitinger, F.} (\bibinfo{year}{2020}).
\newblock \bibinfo{title}{Artifacts for {Detecting} {Timestamp} {Manipulation} in {NTFS} on {Windows} and {Their} {Reliability}}.
\newblock {\it \bibinfo{journal}{Forensic Science International: Digital Investigation}\/},  {\it \bibinfo{volume}{32}\/}, \bibinfo{pages}{300920}. \DOIprefix\doi{10.1016/j.fsidi.2020.300920}.
\bibitem[{Patterson \& Hargreaves(2012)}]{patterson2012potential}
\bibinfo{author}{Patterson, J.}, \& \bibinfo{author}{Hargreaves, C.~J.} (\bibinfo{year}{2012}).
\newblock \bibinfo{title}{The {Potential} for cross-drive analysis using automated digital forensic timelines}.
\newblock \bibinfo{howpublished}{\url{https://dspace.lib.cranfield.ac.uk/handle/1826/8088}}.
\newblock \bibinfo{note}{Accepted: 2014-01-23T05:01:12Z}.
\bibitem[{Quick \& Choo(2014)}]{quick2014impacts}
\bibinfo{author}{Quick, D.}, \& \bibinfo{author}{Choo, K.-K.~R.} (\bibinfo{year}{2014}).
\newblock \bibinfo{title}{Impacts of increasing volume of digital forensic data: A survey and future research challenges}.
\newblock {\it \bibinfo{journal}{Digital Investigation}\/},  {\it \bibinfo{volume}{11}\/}, \bibinfo{pages}{273--294}.
\bibitem[{Raghavan \& Saran(2013)}]{raghavan2013unitime}
\bibinfo{author}{Raghavan, S.}, \& \bibinfo{author}{Saran, H.} (\bibinfo{year}{2013}).
\newblock \bibinfo{title}{Unitime: Timestamp interpretation engine for developing unified timelines}.
\newblock In {\it \bibinfo{booktitle}{2013 8th International Workshop on Systematic Approaches to Digital Forensics Engineering (SADFE)}\/} (pp. \bibinfo{pages}{1--7}).
\newblock \bibinfo{organization}{IEEE}.
\bibitem[{Raju et~al.(2017)Raju, Gosala \& Geethakumari}]{raju2017closer}
\bibinfo{author}{Raju, B.~K.}, \bibinfo{author}{Gosala, N.~B.}, \& \bibinfo{author}{Geethakumari, G.} (\bibinfo{year}{2017}).
\newblock \bibinfo{title}{Closer: applying aggregation for effective event reconstruction of cloud service logs}.
\newblock In {\it \bibinfo{booktitle}{Proceedings of the 11th International Conference on Ubiquitous Information Management and Communication}\/} (pp. \bibinfo{pages}{1--8}).
\bibitem[{Reddy \& Venter(2013)}]{reddy2013architecture}
\bibinfo{author}{Reddy, K.}, \& \bibinfo{author}{Venter, H.~S.} (\bibinfo{year}{2013}).
\newblock \bibinfo{title}{The architecture of a digital forensic readiness management system}.
\newblock {\it \bibinfo{journal}{Computers \& security}\/},  {\it \bibinfo{volume}{32}\/}, \bibinfo{pages}{73--89}. \DOIprefix\doi{10.1016/j.cose.2012.09.008}.
\bibitem[{Ribaux(2014)}]{ribaux2014police}
\bibinfo{author}{Ribaux, O.} (\bibinfo{year}{2014}).
\newblock {\it \bibinfo{title}{Police scientifique: le renseignement par la trace}\/}.
\newblock Sciences forensiques.
\newblock \bibinfo{address}{Lausanne}: \bibinfo{publisher}{Presses polytechniques et universitaires romandes}.
\bibitem[{Ribaux(2023)}]{ribaux2023police}
\bibinfo{author}{Ribaux, O.} (\bibinfo{year}{2023}).
\newblock {\it \bibinfo{title}{De la police scientifique {\`a} la tra{\c{c}}ologie: le renseignement par la trace}\/}.
\newblock \bibinfo{publisher}{EPFL Press}.
\bibitem[{Rivera-Ortiz \& Pasquale(2019)}]{rivera2019towards}
\bibinfo{author}{Rivera-Ortiz, F.}, \& \bibinfo{author}{Pasquale, L.} (\bibinfo{year}{2019}).
\newblock \bibinfo{title}{Towards automated logging for forensic-ready software systems}.
\newblock In {\it \bibinfo{booktitle}{2019 IEEE 27th International Requirements Engineering Conference Workshops (REW)}\/} (pp. \bibinfo{pages}{157--163}).
\newblock \bibinfo{organization}{IEEE}.
\newblock \DOIprefix\doi{10.1109/REW.2019.00033}.
\bibitem[{Roux et~al.(2022)Roux, Bucht, Crispino, De~Forest, Lennard, Margot, Miranda, NicDaeid, Ribaux, Ross \& Willis}]{roux2022sydney}
\bibinfo{author}{Roux, C.}, \bibinfo{author}{Bucht, R.}, \bibinfo{author}{Crispino, F.}, \bibinfo{author}{De~Forest, P.}, \bibinfo{author}{Lennard, C.}, \bibinfo{author}{Margot, P.}, \bibinfo{author}{Miranda, M.~D.}, \bibinfo{author}{NicDaeid, N.}, \bibinfo{author}{Ribaux, O.}, \bibinfo{author}{Ross, A.}, \& \bibinfo{author}{Willis, S.} (\bibinfo{year}{2022}).
\newblock \bibinfo{title}{The {Sydney} declaration – {Revisiting} the essence of forensic science through its fundamental principles}.
\newblock {\it \bibinfo{journal}{Forensic Science International}\/},  {\it \bibinfo{volume}{332}\/}, \bibinfo{pages}{111182}. \DOIprefix\doi{10.1016/j.forsciint.2022.111182}.
\bibitem[{Sachowski(2019)}]{sachowski2019implementing}
\bibinfo{author}{Sachowski, J.} (\bibinfo{year}{2019}).
\newblock {\it \bibinfo{title}{Implementing digital forensic readiness: From reactive to proactive process}\/}.
\newblock \bibinfo{publisher}{CRC Press}.
\newblock \DOIprefix\doi{10.1016/C2015-0-00701-8}.
\bibitem[{Sandvik et~al.(2021)Sandvik, Franke \& {\AA}rnes}]{sandvik2021towards}
\bibinfo{author}{Sandvik, J.-P.}, \bibinfo{author}{Franke, K.}, \& \bibinfo{author}{{\AA}rnes, A.} (\bibinfo{year}{2021}).
\newblock \bibinfo{title}{Towards a generic approach of quantifying evidence volatility in resource constrained devices}.
\newblock {\it \bibinfo{journal}{Digital Forensic Investigation of Internet of Things (IoT) Devices}\/},  (pp. \bibinfo{pages}{21--45}).
\bibitem[{Sandvik \& Årnes(2018)}]{sandvik2018reliability}
\bibinfo{author}{Sandvik, J.-P.}, \& \bibinfo{author}{Årnes, A.} (\bibinfo{year}{2018}).
\newblock \bibinfo{title}{The reliability of clocks as digital evidence under low voltage conditions}.
\newblock {\it \bibinfo{journal}{Digital Investigation}\/},  {\it \bibinfo{volume}{24}\/}, \bibinfo{pages}{S10--S17}. \DOIprefix\doi{10.1016/j.diin.2018.01.003}.
\bibitem[{Scanlon et~al.(2023)Scanlon, Breitinger, Hargreaves, Hilgert \& Sheppard}]{scanlon2023chatgpt}
\bibinfo{author}{Scanlon, M.}, \bibinfo{author}{Breitinger, F.}, \bibinfo{author}{Hargreaves, C.}, \bibinfo{author}{Hilgert, J.-N.}, \& \bibinfo{author}{Sheppard, J.} (\bibinfo{year}{2023}).
\newblock \bibinfo{title}{{ChatGPT} for digital forensic investigation: The good, the bad, and the unknown}.
\newblock {\it \bibinfo{journal}{Forensic Science International: Digital Investigation}\/},  {\it \bibinfo{volume}{46}\/}, \bibinfo{pages}{301609}.
\bibitem[{Schatz et~al.(2006)Schatz, Mohay \& Clark}]{schatz2006correlation}
\bibinfo{author}{Schatz, B.}, \bibinfo{author}{Mohay, G.}, \& \bibinfo{author}{Clark, A.} (\bibinfo{year}{2006}).
\newblock \bibinfo{title}{A correlation method for establishing provenance of timestamps in digital evidence}.
\newblock {\it \bibinfo{journal}{Digital Investigation}\/},  {\it \bibinfo{volume}{3}\/}, \bibinfo{pages}{98--107}. \DOIprefix\doi{10.1016/j.diin.2006.06.009}.
\bibitem[{Schneider et~al.(2022)Schneider, Düsel, Lorch, Drafz \& Freiling}]{schneider2022prudent}
\bibinfo{author}{Schneider, J.}, \bibinfo{author}{Düsel, L.}, \bibinfo{author}{Lorch, B.}, \bibinfo{author}{Drafz, J.}, \& \bibinfo{author}{Freiling, F.} (\bibinfo{year}{2022}).
\newblock \bibinfo{title}{Prudent design principles for digital tampering experiments}.
\newblock {\it \bibinfo{journal}{Forensic Science International: Digital Investigation}\/},  {\it \bibinfo{volume}{40}\/}, \bibinfo{pages}{301334}. \DOIprefix\doi{10.1016/j.fsidi.2022.301334}.
\bibitem[{Schneider et~al.(2024)Schneider, Eichhorn, Dreier \& Hargreaves}]{schneider2024applying}
\bibinfo{author}{Schneider, J.}, \bibinfo{author}{Eichhorn, M.}, \bibinfo{author}{Dreier, L.~M.}, \& \bibinfo{author}{Hargreaves, C.} (\bibinfo{year}{2024}).
\newblock \bibinfo{title}{Applying digital stratigraphy to the problem of recycled storage media}.
\newblock {\it \bibinfo{journal}{Forensic Science International: Digital Investigation}\/},  {\it \bibinfo{volume}{49}\/}, \bibinfo{pages}{301761}.
\bibitem[{Schneider et~al.(2020)Schneider, Wolf \& Freiling}]{schneider2020tampering}
\bibinfo{author}{Schneider, J.}, \bibinfo{author}{Wolf, J.}, \& \bibinfo{author}{Freiling, F.} (\bibinfo{year}{2020}).
\newblock \bibinfo{title}{Tampering with {Digital} {Evidence} is {Hard}: {The} {Case} of {Main} {Memory} {Images}}.
\newblock {\it \bibinfo{journal}{Forensic Science International: Digital Investigation}\/},  {\it \bibinfo{volume}{32}\/}, \bibinfo{pages}{300924}. \DOIprefix\doi{10.1016/j.fsidi.2020.300924}.
\bibitem[{Schuster(2007)}]{schuster2007introducing}
\bibinfo{author}{Schuster, A.} (\bibinfo{year}{2007}).
\newblock \bibinfo{title}{Introducing the microsoft vista event log file format}.
\newblock {\it \bibinfo{journal}{Digital Investigation}\/},  {\it \bibinfo{volume}{4}\/}, \bibinfo{pages}{65--72}.
\bibitem[{Silalahi et~al.(2023{\natexlab{a}})Silalahi, Ahmad \& Studiawan}]{silalahi2023dfler}
\bibinfo{author}{Silalahi, S.}, \bibinfo{author}{Ahmad, T.}, \& \bibinfo{author}{Studiawan, H.} (\bibinfo{year}{2023}{\natexlab{a}}).
\newblock \bibinfo{title}{Dfler: Drone flight log entity recognizer to support forensic investigation on drone device}.
\newblock {\it \bibinfo{journal}{Software Impacts}\/},  {\it \bibinfo{volume}{15}\/}, \bibinfo{pages}{100457}. \DOIprefix\doi{10.1016/j.simpa.2022.100457}.
\bibitem[{Silalahi et~al.(2023{\natexlab{b}})Silalahi, Ahmad \& Studiawan}]{silalahi2023transformer-ner}
\bibinfo{author}{Silalahi, S.}, \bibinfo{author}{Ahmad, T.}, \& \bibinfo{author}{Studiawan, H.} (\bibinfo{year}{2023}{\natexlab{b}}).
\newblock \bibinfo{title}{Transformer-based named entity recognition on drone flight logs to support forensic investigation}.
\newblock {\it \bibinfo{journal}{IEEE Access}\/},  {\it \bibinfo{volume}{11}\/}, \bibinfo{pages}{3257--3274}. \DOIprefix\doi{10.1109/ACCESS.2023.3234605}.
\bibitem[{Silalahi et~al.(2023{\natexlab{c}})Silalahi, Ahmad \& Studiawan}]{silalahi2023transformer-sentiment}
\bibinfo{author}{Silalahi, S.}, \bibinfo{author}{Ahmad, T.}, \& \bibinfo{author}{Studiawan, H.} (\bibinfo{year}{2023}{\natexlab{c}}).
\newblock \bibinfo{title}{Transformer-based sentiment analysis for anomaly detection on drone forensic timeline}.
\newblock In {\it \bibinfo{booktitle}{2023 11th International Symposium on Digital Forensics and Security (ISDFS)}\/} (pp. \bibinfo{pages}{1--6}).
\newblock \bibinfo{organization}{IEEE}.
\newblock \DOIprefix\doi{ISDFS58141.2023.10131749}.
\bibitem[{Soltani et~al.(2019)Soltani, Hosseini~Seno \& sadoghi yazdi}]{soltani2019event}
\bibinfo{author}{Soltani, S.}, \bibinfo{author}{Hosseini~Seno, S.~A.}, \& \bibinfo{author}{sadoghi yazdi, H.} (\bibinfo{year}{2019}).
\newblock \bibinfo{title}{Event {Reconstruction} using {Temporal} {Pattern} of {File} {System} {Modification}}.
\newblock {\it \bibinfo{journal}{IET Information Security}\/},  {\it \bibinfo{volume}{13}\/}. \DOIprefix\doi{10.1049/iet-ifs.2018.5209}.
\bibitem[{Soltani \& Seno(2017)}]{soltani2017survey}
\bibinfo{author}{Soltani, S.}, \& \bibinfo{author}{Seno, S. A.~H.} (\bibinfo{year}{2017}).
\newblock \bibinfo{title}{A survey on digital evidence collection and analysis}.
\newblock In {\it \bibinfo{booktitle}{2017 7th International Conference on Computer and Knowledge Engineering (ICCKE)}\/} (pp. \bibinfo{pages}{247--253}).
\newblock \bibinfo{organization}{IEEE}.
\bibitem[{Soltani \& Seno(2019)}]{soltani2019formal}
\bibinfo{author}{Soltani, S.}, \& \bibinfo{author}{Seno, S. A.~H.} (\bibinfo{year}{2019}).
\newblock \bibinfo{title}{A formal model for event reconstruction in digital forensic investigation}.
\newblock {\it \bibinfo{journal}{Digital Investigation}\/},  {\it \bibinfo{volume}{30}\/}, \bibinfo{pages}{148--160}. \DOIprefix\doi{10.1016/j.diin.2019.07.006}.
\bibitem[{Song et~al.(2016)Song, Cao \& Wang}]{song2016cleaning}
\bibinfo{author}{Song, S.}, \bibinfo{author}{Cao, Y.}, \& \bibinfo{author}{Wang, J.} (\bibinfo{year}{2016}).
\newblock \bibinfo{title}{Cleaning timestamps with temporal constraints}.
\newblock {\it \bibinfo{journal}{Proceedings of the VLDB Endowment}\/},  {\it \bibinfo{volume}{9}\/}, \bibinfo{pages}{708--719}.
\bibitem[{Spichiger \& Adelstein(2025)}]{SPICHIGER2025301867}
\bibinfo{author}{Spichiger, H.}, \& \bibinfo{author}{Adelstein, F.} (\bibinfo{year}{2025}).
\newblock \bibinfo{title}{Preserving meaning of evidence from evolving systems}.
\newblock {\it \bibinfo{journal}{Forensic Science International: Digital Investigation}\/},  {\it \bibinfo{volume}{52}\/}, \bibinfo{pages}{301867}. \URLprefix \url{https://www.sciencedirect.com/science/article/pii/S266628172500006X}. \DOIprefix\doi{https://doi.org/10.1016/j.fsidi.2025.301867}.
\newblock \bibinfo{note}{DFRWS EU 2025 - Selected Papers from the 12th Annual Digital Forensics Research Conference Europe}.
\bibitem[{Sreya et~al.(2023)Sreya, Wadhwa et~al.}]{sreya2023enhancing}
\bibinfo{author}{Sreya, E.}, \bibinfo{author}{Wadhwa, M.} et~al. (\bibinfo{year}{2023}).
\newblock \bibinfo{title}{Enhancing digital investigation: Leveraging chatgpt for evidence identification and analysis in digital forensics}.
\newblock In {\it \bibinfo{booktitle}{2023 International Conference on Computing, Communication, and Intelligent Systems (ICCCIS)}\/} (pp. \bibinfo{pages}{733--738}).
\newblock \bibinfo{organization}{IEEE}.
\bibitem[{Stevens(2004)}]{stevens2004unification}
\bibinfo{author}{Stevens, M.~W.} (\bibinfo{year}{2004}).
\newblock \bibinfo{title}{Unification of relative time frames for digital forensics}.
\newblock {\it \bibinfo{journal}{Digital Investigation}\/},  {\it \bibinfo{volume}{1}\/}, \bibinfo{pages}{225--239}. \DOIprefix\doi{10.1016/j.diin.2004.07.003}.
\bibitem[{Studiawan(2023)}]{studiawan2023event}
\bibinfo{author}{Studiawan, H.} (\bibinfo{year}{2023}).
\newblock \bibinfo{title}{Event abstration in a forensic timeline}.
\newblock In {\it \bibinfo{booktitle}{International Conference for Information and Communication Technologies}\/} (pp. \bibinfo{pages}{119--129}).
\newblock \bibinfo{organization}{Springer}.
\bibitem[{Studiawan et~al.(2023)Studiawan, Hasan \& Pratomo}]{studiawan2023rule}
\bibinfo{author}{Studiawan, H.}, \bibinfo{author}{Hasan, M.~F.}, \& \bibinfo{author}{Pratomo, B.~A.} (\bibinfo{year}{2023}).
\newblock \bibinfo{title}{Rule-based entity recognition for forensic timeline}.
\newblock In {\it \bibinfo{booktitle}{2023 Conference on Information Communications Technology and Society (ICTAS)}\/} (pp. \bibinfo{pages}{1--6}).
\newblock \bibinfo{organization}{IEEE}.
\bibitem[{Studiawan et~al.(2017)Studiawan, Payne \& Sohel}]{studiawan2017graph}
\bibinfo{author}{Studiawan, H.}, \bibinfo{author}{Payne, C.}, \& \bibinfo{author}{Sohel, F.} (\bibinfo{year}{2017}).
\newblock \bibinfo{title}{Graph clustering and anomaly detection of access control log for forensic purposes}.
\newblock {\it \bibinfo{journal}{Digital Investigation}\/},  {\it \bibinfo{volume}{21}\/}, \bibinfo{pages}{76--87}.
\bibitem[{Studiawan \& Sohel(2021)}]{studiawan2021anomaly}
\bibinfo{author}{Studiawan, H.}, \& \bibinfo{author}{Sohel, F.} (\bibinfo{year}{2021}).
\newblock \bibinfo{title}{Anomaly detection in a forensic timeline with deep autoencoders}.
\newblock {\it \bibinfo{journal}{Journal of Information Security and Applications}\/},  {\it \bibinfo{volume}{63}\/}, \bibinfo{pages}{103002}.
\bibitem[{Studiawan et~al.(2019)Studiawan, Sohel \& Payne}]{studiawan2019survey}
\bibinfo{author}{Studiawan, H.}, \bibinfo{author}{Sohel, F.}, \& \bibinfo{author}{Payne, C.} (\bibinfo{year}{2019}).
\newblock \bibinfo{title}{A survey on forensic investigation of operating system logs}.
\newblock {\it \bibinfo{journal}{Digital Investigation}\/},  {\it \bibinfo{volume}{29}\/}, \bibinfo{pages}{1--20}. \DOIprefix\doi{10.1016/j.diin.2019.02.005}.
\bibitem[{Studiawan et~al.(2020{\natexlab{a}})Studiawan, Sohel \& Payne}]{studiawan2020automatic}
\bibinfo{author}{Studiawan, H.}, \bibinfo{author}{Sohel, F.}, \& \bibinfo{author}{Payne, C.} (\bibinfo{year}{2020}{\natexlab{a}}).
\newblock \bibinfo{title}{Automatic event log abstraction to support forensic investigation}.
\newblock In {\it \bibinfo{booktitle}{Proceedings of the Australasian Computer Science Week Multiconference}\/} (pp. \bibinfo{pages}{1--9}).
\bibitem[{Studiawan et~al.(2020{\natexlab{b}})Studiawan, Sohel \& Payne}]{studiawan2020sentiment}
\bibinfo{author}{Studiawan, H.}, \bibinfo{author}{Sohel, F.}, \& \bibinfo{author}{Payne, C.} (\bibinfo{year}{2020}{\natexlab{b}}).
\newblock \bibinfo{title}{Sentiment analysis in a forensic timeline with deep learning}.
\newblock {\it \bibinfo{journal}{IEEE Access}\/},  {\it \bibinfo{volume}{8}\/}, \bibinfo{pages}{60664--60675}. \DOIprefix\doi{10.1109/ACCESS.2020.2983435}.
\bibitem[{Thierry \& M{\"u}ller(2022)}]{thierry2022systematic}
\bibinfo{author}{Thierry, A.}, \& \bibinfo{author}{M{\"u}ller, T.} (\bibinfo{year}{2022}).
\newblock \bibinfo{title}{A systematic approach to understanding {MACB} timestamps on {Unix-like} systems}.
\newblock {\it \bibinfo{journal}{Forensic Science International: Digital Investigation}\/},  {\it \bibinfo{volume}{40}\/}, \bibinfo{pages}{301338}.
\bibitem[{Turnbull \& Randhawa(2015)}]{turnbull2015automated}
\bibinfo{author}{Turnbull, B.}, \& \bibinfo{author}{Randhawa, S.} (\bibinfo{year}{2015}).
\newblock \bibinfo{title}{Automated event and social network extraction from digital evidence sources with ontological mapping}.
\newblock {\it \bibinfo{journal}{Digital Investigation}\/},  {\it \bibinfo{volume}{13}\/}, \bibinfo{pages}{94--106}. \DOIprefix\doi{10.1016/j.diin.2015.04.004}.
\bibitem[{Vanini et~al.(2023)Vanini, Breitinger \& Hargreaves}]{vanini2023presentation}
\bibinfo{author}{Vanini, C.}, \bibinfo{author}{Breitinger, F.}, \& \bibinfo{author}{Hargreaves, C.} (\bibinfo{year}{2023}).
\newblock \bibinfo{title}{A discussion of sources and quality/reliability of events for timelines}.
\newblock \bibinfo{howpublished}{Presentation at the Digital Forensics Research Conference 2023 (Bonn, Germany)}.
\bibitem[{Vanini et~al.(2024{\natexlab{a}})Vanini, Gruber, Hargreaves, Benenson, Freiling \& Breitinger}]{vanini2024strategies}
\bibinfo{author}{Vanini, C.}, \bibinfo{author}{Gruber, J.}, \bibinfo{author}{Hargreaves, C.}, \bibinfo{author}{Benenson, Z.}, \bibinfo{author}{Freiling, F.}, \& \bibinfo{author}{Breitinger, F.} (\bibinfo{year}{2024}{\natexlab{a}}).
\newblock \bibinfo{title}{Strategies and challenges of timestamp tampering for improved digital forensic event reconstruction (extended version)}.
\newblock {\it \bibinfo{journal}{arXiv preprint arXiv:2501.00175}\/}, .
\bibitem[{Vanini et~al.(2024{\natexlab{b}})Vanini, Hargreaves, van Beek \& Breitinger}]{vanini2024clock}
\bibinfo{author}{Vanini, C.}, \bibinfo{author}{Hargreaves, C.~J.}, \bibinfo{author}{van Beek, H.}, \& \bibinfo{author}{Breitinger, F.} (\bibinfo{year}{2024}{\natexlab{b}}).
\newblock \bibinfo{title}{Was the clock correct? {Exploring} timestamp interpretation through time anchors for digital forensic event reconstruction}.
\newblock {\it \bibinfo{journal}{Forensic Science International: Digital Investigation}\/},  {\it \bibinfo{volume}{49}\/}, \bibinfo{pages}{301759}. \DOIprefix\doi{10.1016/j.fsidi.2024.301759}.
\bibitem[{{VMware}(2008)}]{vmware2008timekeeping}
\bibinfo{author}{{VMware}} (\bibinfo{year}{2008}).
\newblock \bibinfo{title}{Timekeeping in {VMware} virtual machines}.
\newblock \bibinfo{howpublished}{https://www.cse.psu.edu/~buu1/teaching/spring06/papers/vmware-timing.pdf}.
\bibitem[{Weijters \& van~der Aalst(2001)}]{weijters2001process}
\bibinfo{author}{Weijters, A.}, \& \bibinfo{author}{van~der Aalst, W.~M.} (\bibinfo{year}{2001}).
\newblock \bibinfo{title}{Process mining: {Discovering} workflow models from event-based data}.
\newblock In {\it \bibinfo{booktitle}{Belgium-Netherlands Conf. on Artificial Intelligence}\/}.
\bibitem[{Willassen(2008{\natexlab{a}})}]{willassen2008hypothesis}
\bibinfo{author}{Willassen, S.} (\bibinfo{year}{2008}{\natexlab{a}}).
\newblock \bibinfo{title}{Hypothesis-based investigation of digital timestamps}.
\newblock In \bibinfo{editor}{I.~Ray}, \& \bibinfo{editor}{S.~Shenoi} (Eds.), {\it \bibinfo{booktitle}{Advances in Digital Forensics IV IFIP — The International Federation for Information Processing}\/} (pp. \bibinfo{pages}{75--86}).
\newblock \bibinfo{address}{Boston, MA}: \bibinfo{publisher}{Springer US} volume \bibinfo{volume}{285}.
\newblock \DOIprefix\doi{10.1007/978-0-387-84927-0_7}.
\bibitem[{Willassen(2008{\natexlab{b}})}]{willassen2008finding}
\bibinfo{author}{Willassen, S.~Y.} (\bibinfo{year}{2008}{\natexlab{b}}).
\newblock \bibinfo{title}{Finding evidence of antedating in digital investigations}.
\newblock In {\it \bibinfo{booktitle}{2008 {Third} {International} {Conference} on {Availability}, {Reliability} and {Security}}\/} (pp. \bibinfo{pages}{26--32}).
\newblock \DOIprefix\doi{10.1109/ARES.2008.149}.
\bibitem[{Willassen(2008{\natexlab{c}})}]{willassen2008timestamp}
\bibinfo{author}{Willassen, S.~Y.} (\bibinfo{year}{2008}{\natexlab{c}}).
\newblock \bibinfo{title}{Timestamp evidence correlation by model based clock hypothesis testing}.
\newblock In {\it \bibinfo{booktitle}{Proceedings of the 1st International Conference on Forensic Applications and Techniques in Telecommunications, Information, and Multimedia and Workshop}\/} e-{Forensics} '08 (pp. \bibinfo{pages}{1--6}).
\newblock \bibinfo{address}{Brussels, BEL}: \bibinfo{publisher}{ICST (Institute for Computer Sciences, Social-Informatics and Telecommunications Engineering)}.
\bibitem[{Xu \& Xu(2022)}]{xu2022visualizing}
\bibinfo{author}{Xu, W.}, \& \bibinfo{author}{Xu, D.} (\bibinfo{year}{2022}).
\newblock \bibinfo{title}{Visualizing and reasoning about presentable digital forensic evidence with knowledge graphs}.
\newblock In {\it \bibinfo{booktitle}{2022 19th Annual International Conference on Privacy, Security \& Trust (PST)}\/} (pp. \bibinfo{pages}{1--10}).
\newblock \bibinfo{organization}{IEEE}.

\end{thebibliography}




\end{document}